\documentclass{egpubl}
\usepackage{egsr2024}

\CGFccby
\usepackage[T1]{fontenc}
\usepackage{dfadobe}
\usepackage{amsmath}
\usepackage{tabularx}
\usepackage[export]{adjustbox}
\usepackage{appendix}
\usepackage{pgfplots}
\usepackage{pgfplotstable} % For \pgfplotstableread
\pgfplotsset{compat=1.17} % You can change this to your pgfplots version
\usepackage{rotating}
\usepackage{float}

\biberVersion
\BibtexOrBiblatex
\usepackage[backend=biber,bibstyle=EG,citestyle=alphabetic,backref=true]{biblatex} 
\electronicVersion
\PrintedOrElectronic

\usepackage{graphicx}
\usepackage{egweblnk} 
\title[Neural Appearance Model for Cloth Rendering]%
      {Neural Appearance Model for Cloth Rendering}

\author[G. Y. Soh \& Z. Montazeri]
{\parbox{\textwidth}{\centering G.\,Y. Soh\orcid{0009-0003-2536-4361}
        and Z. Montazeri\orcid{0000-0003-0398-3105} 
        }
        \\
{\parbox{\textwidth}{\centering University of Manchester, United Kingdom\\
       }
}
}
\usepackage{booktabs} % For prettier tables

\begin{document}

\teaser{
 \includegraphics[width=\linewidth]{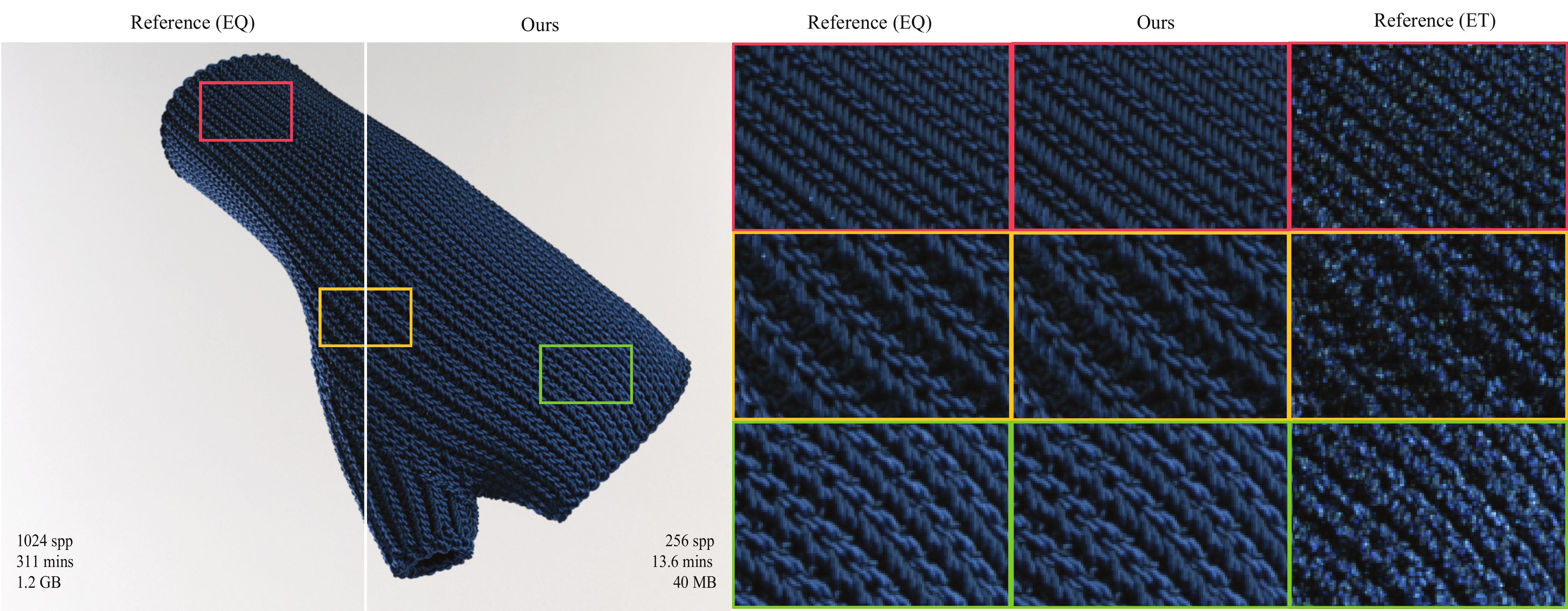}
 \centering
  \caption{In this scene, we compare our neural aggregated model against the reference, which uses explicit fiber-based geometry and scattering model. The left shows the renderings side by side with equal quality (EQ) of the knitted fleece which is 23 times slower and requires 300 times more memory than ours. whereas the right shows the close-up crops to show the equal time (ET) is much noisier than ours.} %\\[2em]}
\label{fig:teaser}
}

\maketitle
%-------------------------------------------------------------------------
\begin{abstract}
The realistic rendering of woven and knitted fabrics has posed significant challenges throughout many years. Previously, fiber-based micro-appearance models have achieved considerable success in attaining high levels of realism. However, rendering such models remains complex due to the intricate internal scatterings of hundreds of fibers within a yarn, requiring vast amounts of memory and time to render. In this paper, we introduce a new framework to capture aggregated appearance by tracing many light paths through the underlying fiber geometry. We then employ lightweight neural networks to accurately model the aggregated BSDF, which allows for the precise modeling of a diverse array of materials while offering substantial improvements in speed and reductions in memory. Furthermore, we introduce a novel importance sampling scheme to further speed up the rate of convergence. We validate the efficacy and versatility of our framework through comparisons with preceding fiber-based shading models as well as the most recent yarn-based model.

%-------------------------------------------------------------------------
%  ACM CCS 1998
%  (see https://www.acm.org/publications/computing-classification-system/1998)
% \begin{classification} % according to https://www.acm.org/publications/computing-classification-system/1998
% \CCScat{Computer Graphics}{I.3.3}{Picture/Image Generation}{Line and curve generation}
% \end{classification}
%-------------------------------------------------------------------------
%  ACM CCS 2012
   % (see https://www.acm.org/publications/class-2012)
%The tool at \url{http://dl.acm.org/ccs.cfm} can be used to generate
% CCS codes.
%Example:


\end{abstract}  
%-------------------------------------------------------------------------
\section{Introduction}
Fabrics are important in our everyday lives and their virtual representation has long been a key focus in computer graphics research. Fabrics, with their detailed structure of fibers, plies, and yarns, present a unique hierarchical geometric structure at each aggregation level, offering a wide array of appearances for different cloth types.

The challenge of accurately modeling the detailed geometry and scattering of fabrics has led to the development of different methods, mainly split into curve-based and surface-based shading models. Curve-based models, using the Bidirectional Curve Scattering Distribution Function (BCSDF), aim to explicitly represent individual elements like fibers \cite{khungurn2015matching}, plies \cite{montazeri2020practical} and yarns \cite{Zhu2023yarn}, similar to methods used in hair rendering. However, these models, while accurate, face challenges like long rendering times and high storage needs.

Micro-appearance models focus on representing fabrics at the microscale, detailing each fiber using high-resolution volumes or fiber meshes \cite{zhao2011building}. These models are great at rendering with high detail but are limited in practical use due to their data-intensive nature and their challenges in manipulation and rendering.

In contrast, surface-based models, which depict fabric as a 2D sheet and use specific reflectance models for appearance, are known for being lightweight and user-friendly e.g.\ \cite{sadeghi2013practical}. These models, widely used in the computer graphics industry, can accurately reproduce the overall appearance of fabrics. However, they often fail to capture the fine details necessary for realistic close-up images.

In this paper, we aim to combine the light scattering of a twisted yarn, made up of hundreds of fibers, by simulating the paths of many light rays into the yarn and analyzing their scattering properties. From this analysis, we show that the scattering can be described as three distinct components, and we introduce a new way to model each component using various neural networks and analytical solutions. Additionally, we derive an analytical importance sampling scheme that closely matches the combined scattering distribution. We demonstrate that our model is able to run up to 23 times faster while using up to 600 times less memory when compared to previous fiber-based methods. The memory gain is directly dependent on the fiber count which is often a few hundred. In summary, our main contributions include:
\begin{itemize}
\item We introduce a novel neural framework for modeling the light scattering within a bundle of fibers in the yarns. By dividing the scattering into components, we can efficiently model various types of yarns across a broad range of parameters. Our proposed method runs significantly faster and uses substantially less memory.

\item We further improve on existing neural network approaches by using the channel-wise PReLU activation function to increase performance. We demonstrate its effectiveness by comparing its performance against various model architectures.

\item From our observations, we derive a new analytical fitting of the scattering for importance sampling. We have managed to derive a new observation-based empirical and invertible importance sampling scheme that matches the scattering distribution to further accelerate the rate of convergence.
\end{itemize}

\section{Prior Work}
\emph{Surface-Based Cloth Models} - Cloth rendering has been a subject of extensive research, with various models being developed to achieve a balance between realism and computational efficiency. Traditional models have often depicted cloth as 2D surfaces, utilizing Bidirectional Reflectance Distribution Functions (BRDFs) or Bidirectional Texture Functions (BTFs) to illustrate light-cloth interactions \cite{sattler2003efficient, adabala2003, irawan2012specular, sadeghi2013practical, rainer2019neural, kuznetsov2021neumip, jin2022woven, zhu2023realistic}. While these surface-based models are lightweight and capable of producing high-quality results at mid-to-far distances, they typically lack the fine-grained details necessary for close-up views.

\emph{Micro-appearance Cloth Models} - On the contrary, micro-appearance models have emerged, focusing on the fabric’s micro-geometry down to the fiber level, offering a high fidelity and detail \cite{schroder2011volumetric, zhao2011building, khungurn2015matching, loubet2018, Montazeri2021mechanics, aliaga2017appearance}. However, the high complexity of these models presents a significant challenge in rendering them efficiently. Various precomputation-based rendering methods have been developed to address this, such as the techniques proposed by \cite{zhao2013modular, khungurn2017fast, luan2017fiber} to improve performance and GPU-based methods developed by \cite{Wu2017realtime} for procedurally generated fabrics. Nevertheless, these methods often compromise either on performance or physical accuracy, as well as being difficult to edit and render.

\emph{Aggregation Based Techniques} - In recent years, aggregation-based methodologies have been introduced to the domain of cloth rendering, aiming to model the multiple scatterings of a bundle of fibers implicitly. Montazeri et al. \cite{montazeri2020practical, montazeri2021practical} pioneered an aggregated technique that encapsulates the light scatterings of individual fibers, approximating the overall appearance at the ply level for woven and knitted fabrics, respectively; later followed by the yarn-level extension \cite{khattar2024multiscale}. However, their model, while being fast and practical, is predominantly observation-driven and not efficient for yarns with a high number of plies.

Zhu et al. \cite{Zhu2022fur} advanced the field by proposing a technique to aggregate the scatterings of a bundle of straight fur fibers in a data-driven manner. They then parameterize the aggregated scattering by fitting analytical lobes, followed by the training of a neural network to predict the parameters for the lobes. This model does not accommodate twisted fibers and, being a far-field model, cannot represent yarn-level highlights at close-up views. In a subsequent study \cite{Zhu2023yarn}, the authors introduced an analytical solution designed to accurately approximate the multi-scattering of yarn by utilizing dual scattering theory. However, this model relies heavily on the assumptions inherited from dual scattering theory and also imposes additional assumptions on the fiber shading model. In contrast, our work, while employing similar fiber scattering models and micro-geometry, presents a more generalized model capable of fitting any yarn without necessitating specific assumptions. 

\emph{Neural BRDF Representation} - \cite{chen2020ibrdf, sztrajman2021nbrdf} was one of the firsts to leverage machine learning to represent BRDFs and achieve a high compression rate while preserving the fidelity of the BRDF. In this paper, we improve on Sztrajman et al.'s \cite{sztrajman2021nbrdf} framework to support aggregated yarn scattering, as we demonstrate that using their framework in a naïve manner do not produce optimal results.

% \section{Overview}
% In this section, we aim to give a brief overview of the paper. In \SS\ref{sec:preliminaries}, we share the needed background information such as details about the fiber geometry. \S\ref{sec:yarnshading} looks at the aggregated light scattering and describes it using three separate components. \S\ref{sec:our_neural_approach} explains the structure and implementation details of our neural network. \S\ref{sec:ablation} explores how different neural network structures affect performance. And \S\ref{sec:results} compares the appearance, performance, and memory use of our model against the explicit fiber-based models and previous works.

\section{Preliminaries}
\label{sec:preliminaries}
Every yarn is made up of twisted plies, which in turn consist of hundreds of strands called fibers. In our study, the primary aim is to aggregate a single-ply geometry while explicitly tracing interactions between plies for multi-ply yarn. The arrangement of the fibers around the yarn is characterized by the parameters $N$, $\rho$, and $\alpha$. Here, $N$ represents the number of fibers in the yarn, $\rho$ represents the fiber density, and $\alpha$ describes the twist factor.
\begin{equation}
    \rho = \frac{N r^2}{ R^2}, \quad
    \alpha = \frac{2R\Delta n}{\Delta l},
    \label{eq:alpha}
\end{equation}
where $r$ is the fiber radius, $R$ is the yarn radius, $n$ is the number of revolutions, and $l$ denotes the length along the yarn. Importantly, these parameters are defined such that they are invariant to the yarn's overall scale, allowing us to use our fitted model on all scales of the yarn with the same parameters, without having to re-train the neural networks or re-fit the parameters. The list of all parameters is detailed in Table \ref{table:symb}.

\begin{table}[t]
  % \centering
  \caption{List of important symbols for our neural yarn shading}
     \begin{tabularx}{\linewidth}{l X} 
     \hline\hline
     \textbf{Notation} & \textbf{Definition} \\
     \hline\hline
      $N$ & number of fibers in a yarn \\
      $\alpha$ & twist factor \\
      $\rho$ & fiber density in a yarn \\ [0.5ex]
     \hline
      $C_R$ & attenuation of fiber reflection \\
      $C_{TT}$ & attenuation of fiber transmission \\
      $\beta_{f_R}$ & longitudinal roughness of fiber reflection \\
      $\beta_{f_{TT}}$ & longitudinal roughness of fiber transmission \\
      $\gamma_{f_{TT}}$ & azimuthal roughness of fiber transmission \\
     \hline
      $\omega_i$ & incoming direction relative to yarn frame \\
      $\omega_o$ & outgoing direction relative to yarn frame \\
      $\tilde{\omega_i}$ & incoming direction relative to surface fiber frame \\
      $\tilde{\omega_o}$ & outgoing direction relative to surface fiber frame \\
      $\theta_i$ & incoming longitudinal angle \\
      $\theta_o$ & outgoing longitudinal angle \\
      $\phi_i$ & incoming azimuthal angle \\
      $\phi_o$ & outgoing azimuthal angle \\ [0.5ex]
     \hline
      $S$ & our aggregated yarn scattering function \\
      $S_R$ & our aggregated yarn direct reflection component \\
      $S_T$ & our aggregated yarn direct transmission component \\
      $S_M$ & our aggregated yarn multi-scattering component \\
      $S_f$ & fiber scattering function \\[1ex]
    \hline\hline
     \end{tabularx}     
    \label{table:symb}
\end{table}

\subsection{Fiber Shading Model}
In this paper, our fiber shading model is based on the method by Khungurn et al. \cite{khungurn2015matching}, where fibers are modeled as glass-like tubes and the scatterings are split into Reflection (R) and Transmission (TT) lobes.

\begin{equation}
    S_f(\omega_i, \omega_o) = S_f(\theta_i, \theta_o, \phi_i, \phi_o) = \sum_{k} M_k(\theta_i, \theta_o) N_k(\phi_i, \phi_o),
\end{equation}

where $k=\{R,TT\}$. The incident and outgoing directions $\omega_i$, $\omega_o$ are parameterized into the longitudinal angle $\theta$ and azimuthal angle $\phi$ using the coordinate system defined in Marschner et al. \cite{Marschner2003}. $M$ represents the scattering in the longitudinal plane, and $N$ represents the scattering in the azimuthal plane. They are defined as:
\begin{equation}
\begin{split}
    M_R(\theta_i, \theta_o) &= F_R(\theta_i) \bar{g}(\theta_o; -\theta_i, \beta_{f_R}) \\
    M_{TT}(\theta_i, \theta_o) &=  C_{TT}(1- F_R(\theta_i)) \bar{g}(\theta_o; -\theta_i, \beta_{f_{TT}}) \\
    N_R(\phi_i, \phi_o) &= \frac{1}{2\pi} \\
    N_{TT}(\phi_i, \phi_o) &= f(\phi_o; \phi_i + \pi, \gamma_{f_{TT}}^{-2}),
\end{split}
\end{equation}
where $C_R$, $C_{TT}$ represents the attenuation of each component, $\beta_{f_R}$, $\beta_{f_{TT}}$ represents the longitudinal roughness, and $\gamma_{TT_F}$ represents the azimuthal roughness. $\bar{g}$ is the normalized Gaussian function defined in Khungurn et al.\cite{khungurn2015matching} and $f$ denotes the von Mises distribution. Furthermore, $F_R$ is the Fresnel term and is approximated via Schlick's approximation \cite{schlick1994}:
\begin{equation}
    F_R(\theta_i) = C_R + (1 - C_R)(1 - \cos\theta_i) ^ 5.
\end{equation}

\subsection{Yarn Shading Frame}
In this work, we found it useful to describe the light scattering in terms of two separate shading frames, the yarn shading frame, and the surface fiber shading frame. The yarn shading frame is defined as a traditional anisotropic surface shading frame on the yarn cylinder, with the incident and outgoing directions parameterized with longitudinal angle $\theta$ and azimuthal angle $\phi$. The normal of the frame is aligned with the normal of the cylinder surface, while the tangent of the frame is aligned in the direction of the yarn tangent. We chose this in contrast to existing hair literature, where a longitudinal angle $\theta$ and an azimuthal offset $h$ are used, to make the process of finding the surface fiber shading frame easier. The surface fiber shading frame describes the fiber shading frame on the surface of the yarn, and using the coordinates system of Marschner et al. \cite{Marschner2003} with $\phi=0^\circ$ when pointing towards the surface normal. The frame is rotated around the surface normal due to the fiber twist. In our paper, we denote the directions relative to the yarn shading frame with $\omega_i$ and $\omega_o$, while the directions relative to the surface fiber shading frame as $\tilde{\omega_i}$ and $\tilde{\omega_o}$ which can be defined as:

\begin{equation}
    \tilde{\omega_i} = M \omega_i, \quad
    \tilde{\omega_o} = M \omega_o, \quad
    M = \begin{bmatrix} \cos\tilde{\phi} & -\sin\tilde{\phi} & 0\\ \sin\tilde{\phi} & \cos\tilde{\phi} & 0 \\ 0 & 0 & 1 \end{bmatrix},
\end{equation}
where $\tilde{\phi} = \tan^{-1}(\pi\alpha)$. Derivation of the angle $\tilde{\phi}$ can be found in the appendix below.

\begin{figure*}
    \includegraphics[width=1.0\textwidth]{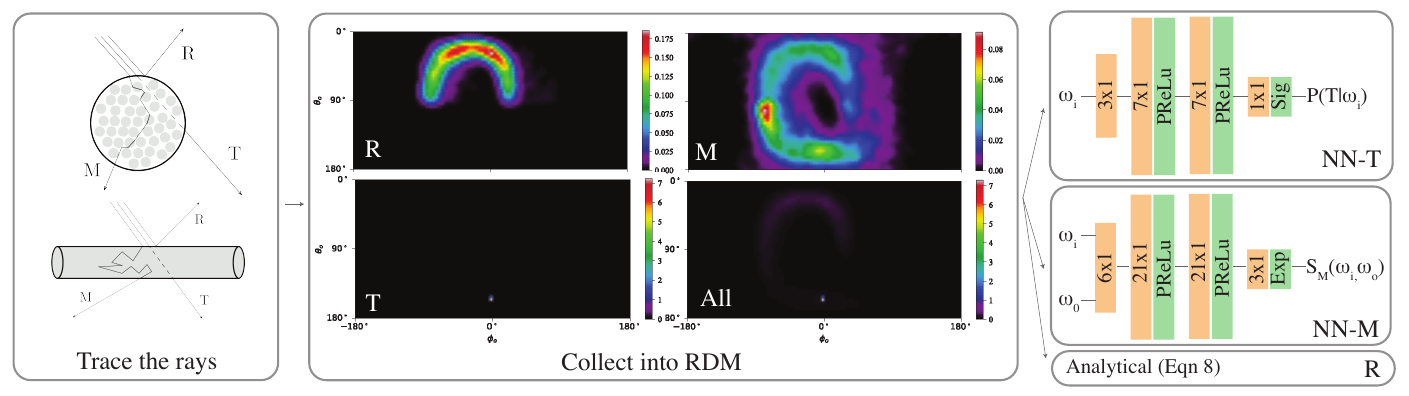}
    \caption{Overview of our pipeline. The first step is to explicitly trace the rays and label them into three components to gather the data (direct transmission T, direct reflection R, multi-scattering M). Next, collect them into Radiance Distribution Maps (RDM). Here, we separate each component of the RDM (T, R, M) to demonstrate the vastly different scales and distributions of each component and visualize for when $\theta_i = 45^\circ$ and $\phi_i = 0^\circ$. Lastly, the networks to learn the T and M components are visualized while R is being computed analytically.}
    \label{fig:pipeline}
\end{figure*}

\begin{figure}[b]
    \centering
    \begin{tabular}{l l}
        All & \includegraphics[width = 0.85\linewidth, trim = 0 6cm 0 6cm, clip, valign=m]{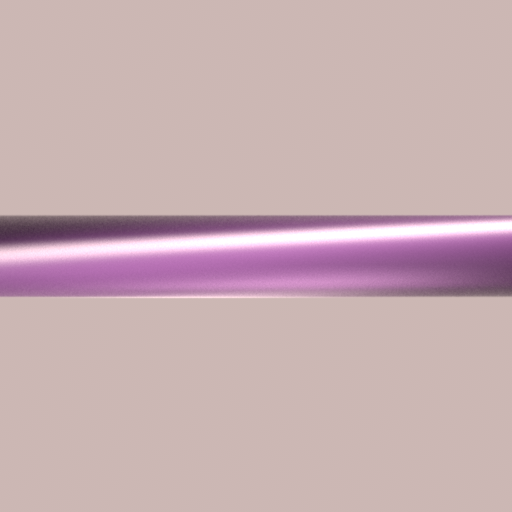} \\ [-0.42cm]
        T & \includegraphics[width = 0.85\linewidth, trim = 0 6cm 0 6cm, clip, valign=m]{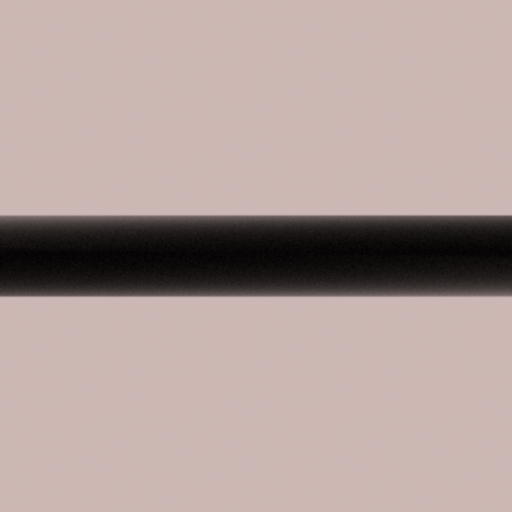} \\ [-0.42cm]
        R & \includegraphics[width = 0.85\linewidth, trim = 0 6cm 0 6cm, clip, valign=m]{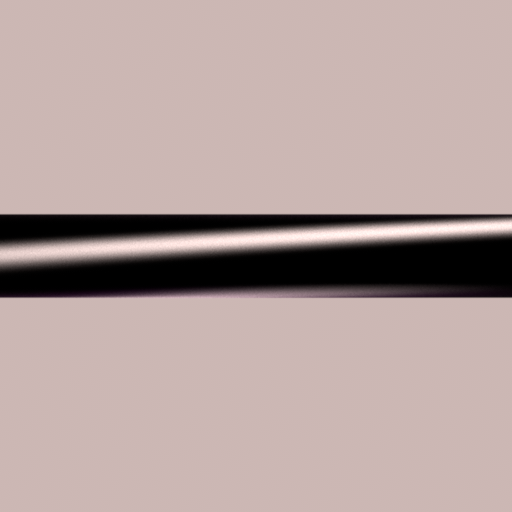} \\ [-0.42cm]
        M & \includegraphics[width = 0.85\linewidth, trim = 0 6cm 0 6cm, clip, valign=m]{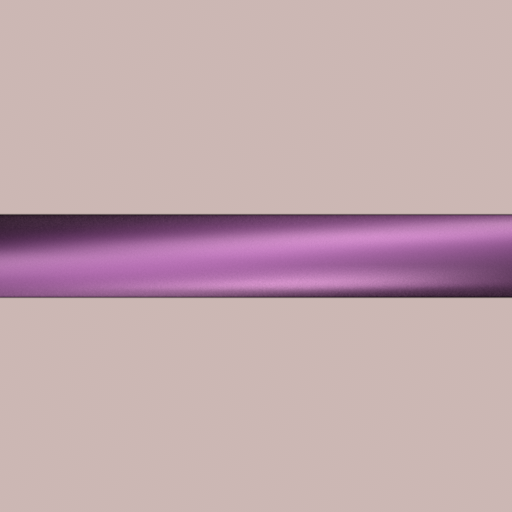} \\
    \end{tabular}
    \caption{The contribution of each component (direct transmission T, direct reflection R, multi-scattering M) to the final appearance of the yarn.}
    \label{fig:components}
\end{figure}
\section{Our Aggregated Shading Model}
\label{sec:yarnshading}
In alignment with the methodology proposed by Zhu et al. \cite{Zhu2022fur}, the aggregation of yarn fibers is achieved by encapsulating them within a closely bounded cylinder. We aggregate the yarn scattering by simulating many light rays into the yarn and recording their exiting radiance and direction to obtain the Radiance Distribution Map (RDM). The RDM is a 4-dimensional map of the exiting radiance for a given $\omega_i$ and $\omega_o$ and is parameterized by $\theta$, $\phi$. Further details on obtaining the RDM are given in \S\ref{sec:multi-scattering}. We then propose a framework to model the RDM by observation and analysis and split the RDM into 3 components to model them individually.

It might intuitively seem advantageous to model the RDM directly by fitting a neural network to it. However, our experiments suggest that this approach is not optimal. Initially, it was observed that at certain incoming angles, specifically at grazing azimuthal angles, a substantial amount of light traverses through the yarn cylinder without interacting with any fibers. In such instances, most of the light is directly transmitted and exhibits Dirac delta distributions, resulting in the corresponding RDM displaying sharp lobes with pronounced peak values. Such distributions pose a significant learning challenge for the neural network due to their high values.

% \begin{figure}[t]
%     \centering
%     \includegraphics[width=\linewidth]{images/illustration/paths.pdf}
%     \caption{Typical light paths for each component. Direct transmission (T) represents paths that do not intersect many fibers within the yarn; direct reflection (R) represents paths that intersect only one fiber before reflecting off the yarn; multi-scattering (M) represents paths that comprise of multiple bounces before leaving the yarn medium.}
%     \label{fig:paths}
% \end{figure}

Furthermore, a considerable fraction of the brightness within the RDM is attributed to the paths characterized by a single bounce. These paths, interacting with a single fiber on the yarn's surface before exiting, create the highlights of the yarn and introduce abrupt alterations in brightness within the RDM. By isolating these paths into a distinct component, we achieved more precise highlights and facilitated the learning process for the neural network regarding the remaining data. Consequently, we introduce the subsequent shading model as a mixture of separate components T, R, and M, corresponding to the Direct Transmission Component, Direct Reflection Component, and Multi-Scattering Component respectively:
\begin{equation}
    S(\omega_i, \omega_o) = S_T(\omega_i, \omega_o) + S_R(\omega_i, \omega_o) + S_M(\omega_i, \omega_o),
\end{equation}
where the T component models the light paths that directly pass through the yarn without intersecting many fibers, the R component models the light paths that hit a single fiber and are reflected away, and the M component models the multiple scattering of light within the yarn before exiting. The components T and M are more complex and hence modelled by a neural network, while the R component can be found analytically. By splitting the shading model into separate components, we can better fit each lobe more accurately, whilst using fewer parameters for the neural network, increasing computational efficiency. The first column in Fig. \ref{fig:pipeline} illustrated the pathways associated with each component, followed by the visualization of the distributions of each component. Fig. \ref{fig:components} visualize the appearance of each component to showcase their contribution individually.

% \begin{figure}[t]
%     \centering
%     \hspace{-25pt}
%     \includegraphics[width=1.05\linewidth]{images/rdm/RDM.pdf}
%     \caption{Typical RDM of a yarn when $\theta_i = 45^\circ$ and $\phi_i = 0^\circ$. Here, we separate each component of the RDM (T, R, M) to demonstrate the vastly different scales and distributions of each component. (direct transmission T, direct reflection R, multi-scattering M)}
%     \label{fig:RDMdiagram}
% \end{figure}

\subsection{Direct Transmission Component}
The Direct Transmission Component of our model represents the fraction of incoming light that directly passes through the yarn without intersecting any fibers, given a specific incident direction $\omega_i$. It becomes particularly prominent in yarn assemblies with lower fiber densities, where a high proportion of light rays pass through directly, resulting in a more translucent appearance. Its influence is also more pronounced at grazing azimuthal angles. Consequently, we incorporate this component into the final scattering function. This component can be mathematically expressed as:
\begin{equation}
    S_T(\omega_i, \omega_o) = P(T|\omega_i) \delta(\omega_i + \omega_o),
\end{equation}
where $\delta(\omega_i + \omega_o)$ is the Dirac delta distribution, which is zero except when $\omega_i = -\omega_o$. The probability $P(T|\omega_i)$ is multiplied with the Dirac delta distribution $\delta(\omega_i + \omega_o)$ to determine the radiance of the transmission component. Instead of fitting $S_T(\omega_i, \omega_o)$ directly with a neural network, we fit $P(T|\omega_i)$. This component is a two-dimensional map and can easily be modelled by a lightweight neural network.

\subsection{Direct Reflection Component}
Since the Direct Reflection Component is the reflection of fibers with a single bounce, this component corresponds to a bright highlight on the yarn surface. Therefore, this component contributes to a sharp change in radiance in the RDM. Hence, it would be beneficial to model this component analytically as opposed to fitting it with a neural network as this would allow us to achieve more accurate highlights, while simultaneously allowing the neural networks to converge at a faster rate with the other parameters. We model this component as a single fiber scattering relative to the surface fiber shading frame on the upper hemisphere of the surface.
\begin{equation}
S_R(\omega_i, \omega_o) =
\begin{cases}
    (1 - P(T|\omega_i)) S_f(\tilde{\omega_i}, \tilde{\omega_o}) & \text{if } \omega_i \cdot \textbf{n} > 0 \\
    0 & \text{otherwise}
\end{cases}
\end{equation}

\subsection{Multi-Scattering Component}
\label{sec:multi-scattering_component}
The Multi-Scattering Component captures the detailed interactions among fibers within the ply or yarn and is represented as $S_M(\omega_i, \omega_o)$ in the scattering function. By using a neural network, we can effectively learn the distribution of these interactions, creating a robust multi-scattering model. This method is especially useful for modeling yarn aggregation to capture the scatterings of more complex yarn geometries, such as twist, a feature that the existing studies overlooked \cite{Zhu2022fur}. For more detailed information and specific details about the network, please refer to \S \ref{sec:multi-scattering}.

\subsection{Importance Sampling}
\begin{figure}[t]
    \centering
    \begin{tabular}{c c}
        Uniform Sampling & Importance Sampling\\ [0.5em]
        \includegraphics[width=0.45\linewidth, trim = 3cm 6cm 3cm 3cm, clip]{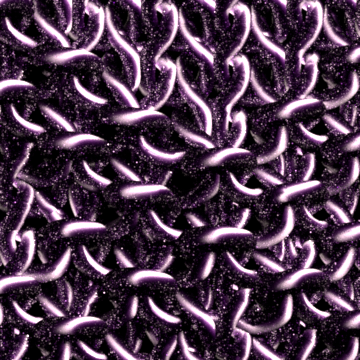} & 
        \includegraphics[width=0.45\linewidth, trim = 3cm 6cm 3cm 3cm, clip]{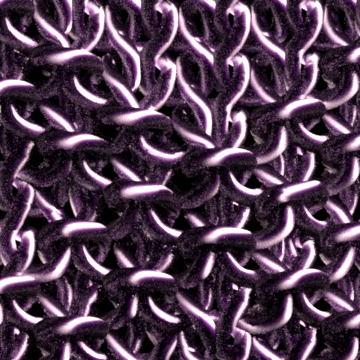} \\
    \end{tabular}
    \caption{Comparison of uniform sampling vs our proposed importance sampling scheme. The images are rendered at 64spp and demonstrate that our importance sampling scheme significantly reduces the variance with less noise.}
    \label{fig:importancevsuniform}
\end{figure}
Given that pieces of cloth are composed of numerous yarns, the inter-reflection amongst the yarns significantly influences the overall visual appearance. It is important to employ an advanced importance sampling scheme to reduce variance as showcased in Fig. \ref{fig:importancevsuniform}. Nonetheless, due to the complexity of light scattering within a yarn when utilizing a neural network in our approach, we are precluded from using the Bidirectional Scattering Distribution Function (BSDF) for importance sampling. Consequently, we chose to fit an invertible analytical approximation of the data to enhance the sampling of the distribution. Sztrajman et al. \cite{sztrajman2021nbrdf} utilized Blinn-Phong lobes to fit the distribution of their Neural BRDFs. However, given that the scattering of light within a yarn does not center around the half angle, the Blinn-Phong lobe is a poor fit for our model. From our observations of the multi-scattering component, we found that the light mainly scatters around the upper half of the cone centered at the fiber tangent at the yarn surface. Thus, we propose the following importance sampling scheme:

\begin{itemize}
    \item  \emph{Sample lobe $k$} - We sample the lobes proportional to the energy of their lobes. Since $S_T(\omega_i, \omega_o)$ is comprised of light passing-through with a probability of $P(T|\omega_i)$, the proportion of energy can be described as $P(T|\omega_i)$ directly. The remaining portion of samples can be split proportionally according to the energy of $S_R(\omega_i, \omega_o)$ and $S_M(\omega_i, \omega_o)$, which can be approximated by a constant $\kappa_R$ which is pre-computed beforehand based on the computed RDM. 
    
    \item \emph{Sample outgoing direction} - For the direct transmission component, we sample in the direction $\omega_o = -\omega_i$ to simulate the light passing through the yarn. The direct reflection is sampled similarly to the fiber’s distribution on the yarn surface. It is done by sampling the longitudinal angles via a normalized Gaussian around $-\theta_i$ with the standard deviation corresponding to the fiber reflection's longitudinal roughness $\beta_{f_R}$, while the azimuthal angle is uniformly distributed on the upper cone in the range $[-\pi/2, \pi/2]$. 
    
    The remaining multi-scattering component is sampled via two lobes which are derived from careful observations of the RDM. The first lobe is comprised of a distribution similar to the direct reflection component but with a different longitudinal and azimuthal roughness. It is defined by a longitudinal Gaussian distribution $g(\theta_i, -\theta_o, \beta_M)$ and azimuthal von Mises distribution $f(\phi_i, 0^\circ, \gamma_M)$, where azimuthal angle is zero at $\textbf{n}$. The second distribution is described by a simple uniform sphere to capture the remaining directions not covered by the first lobe. The two lobes are split with a parameter $\kappa_M$. The parameters $\beta_M$, $\gamma_M$, and $\kappa_M$ are to be fitted beforehand. \\
    \item \emph{Compute the PDF} - The PDF can be described as a mixture of the lobes and can be computed as:
    \begin{equation}
        \text{pdf}(\omega_i, \omega_o) = \sum_{k=\{T, R, M\}} p_k \text{pdf}_k(\tilde{\omega_i}, \tilde{\omega_o})
    \end{equation}
    Here, the PDF for each component is defined as:
    \begin{equation}
    \begin{split}
        \text{pdf}_T(\tilde{\omega_i}, \tilde{\omega_o}) &= \delta(\tilde{\omega_i} + \tilde{\omega_o}) \\ 
        \text{pdf}_R(\tilde{\omega_i}, \tilde{\omega_o}) &= \frac{g(\theta_o; -\theta_i, \beta_{f_R})}{\pi \cos\theta_o} \\
        \text{pdf}_M(\tilde{\omega_i}, \tilde{\omega_o}) &= \frac{\kappa_M g(\theta_o; -\theta_i, \beta_M) f(\phi_o; 0^\circ, \gamma_M^{-2})}{\cos\theta_o} + \frac{1-\kappa_M}{4\pi}
    \end{split}
    \end{equation}
    with their proportions:
    \begin{equation}
    \begin{split}
        p_T &= P(T|wi) \\ 
        p_R &= (1 - p_T) \kappa_R \\
        p_M &= (1 - p_T) (1 - \kappa_R)        
    \end{split}
    \end{equation}
where $g(\theta_o; -\theta_i, \beta)$ represents a Gaussian normalized in the range $[-\pi/2, \pi/2]$ with mean $-\theta_i$ and standard deviation $\beta$. $f(\phi_o; 0^\circ, \gamma_M^{-2})$ represents the von Mises distribution with mean $0^\circ$ and roughness $\gamma_M$.
\end{itemize}

\section{Our Neural Approach}
\label{sec:our_neural_approach}
\begin{figure}[t]
    \centering
    \begin{tabular}{c l}
        \adjustbox{rotate=90,valign=m}{Ours} & \adjustbox{valign=m}{\includegraphics[width = 0.85\linewidth, trim = 0 6cm 0 6cm, clip]{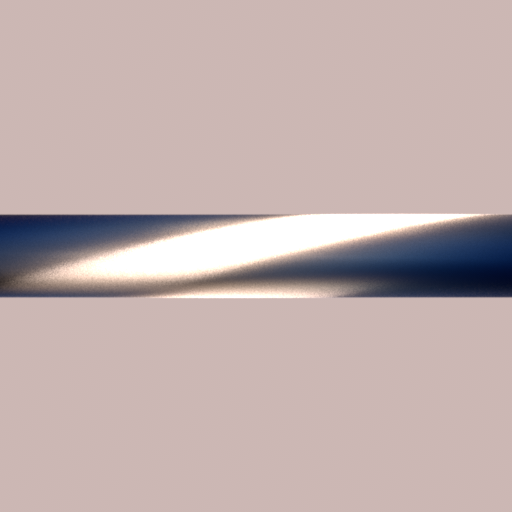}} \\ [35pt]
        \adjustbox{rotate=90,valign=m}{Reference} & \adjustbox{valign=m}{\includegraphics[width = 0.85\linewidth, trim = 0 4.5cm 0 4.5cm, clip]{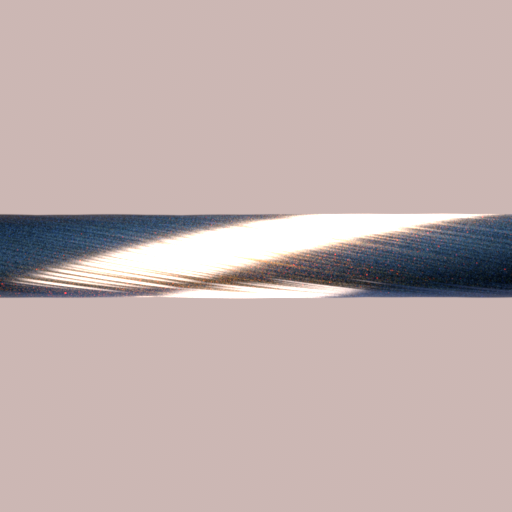}} \\ [35pt]
        \adjustbox{rotate=90,valign=m}{\cite{sztrajman2021nbrdf}} & \adjustbox{valign=m}{\includegraphics[width = 0.85\linewidth, trim = 0 6cm 0 6cm, clip]{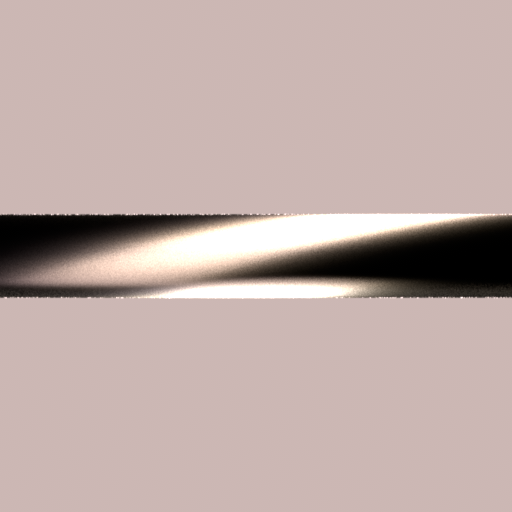}} \\
    \end{tabular}
    \caption{We compare our neural network approach with the naïve approach by contrasting them with the reference. Our method models each component of the RDM as described in \S\ref{sec:yarnshading}, while the naïve approach models the RDM directly using the framework described in \cite{sztrajman2021nbrdf}. Our approach successfully models the reference, including the subtle multi-scatterings, while the latter does not.}
    \label{fig:enter-label}
\end{figure}

Due to the intricate nature of light scattering within a yarn, we opted to model it using neural networks, inspired by the success of Sztrajman et al. \cite{sztrajman2021nbrdf} in accurately and efficiently modeling measured BRDFs. Additionally, this approach provides the generality and flexibility to model various yarn types without making assumptions about the underlying geometry. Beyond the enhancements described in \S\ref{sec:yarnshading} to boost the neural network's performance, we have also refined their base architecture to achieve higher accuracy with minimal runtime costs by employing the channel-wise Parametric Rectified Linear Unit (PReLU) \cite{he2015prelu} activation function instead of the Rectified Linear Unit (ReLU) \cite{nair2010relu} activation presented in their paper. A comparison with this model naïvely is shown in Fig. \ref{fig:enter-label}.

In brief, channel-wise PReLU allows each channel of the input its own learnable parameter, which provides the model with additional flexibility to learn more complex representations without a substantial increase in computational cost and mitigates issues related to the "dying ReLU" problem. The "dying ReLU" problem refers to the phenomenon where neurons in a network become inactive and only output zero during training, essentially ceasing to learn or update and thereby reducing the capacity of the model. This often occurs when a large gradient flows through a ReLU neuron, updating the weights in such a way that the neuron will always output zero. PReLU helps to avoid this issue by maintaining active learning and adapting its negative slope to the learned features of the input data.

Additionally, We chose the channel-wise PReLU activation function over ReLU because it introduces additional trainable parameters for the negative values of ReLU, allowing the neural network more flexibility to overfit with nearly no extra runtime cost, while avoiding the instability of the dying ReLU problem, which is more prevalent in smaller neural networks. Please refer to \S\ref{sec:ablation} for additional details on the performance of various neural network architectures and activation functions.

\subsection{Data Generation}
\label{ssec:data}
To generate data for computing the RDM and preparing the training data for the neural networks, we initially establish our foundational single-ply yarn geometry, as previously detailed in \S\ref{sec:preliminaries}. A bounding cylinder is defined around the yarn, and light rays, each possessing an initial weight of 1, are projected at random directions $\omega_i$, uniformly distributed over a hemisphere, into the yarn. Monte Carlo random walks are subsequently utilized to trace the interactions of each ray with the fibers until it exits the yarn cylinder. For each sample, variables such as the incident angle, outgoing angle, outgoing weight, and the number of bounces (depth) are documented. Our dataset consists of 50-100 million sampled rays that are fully traced for each five yarn materials, with a maximum bounce depth of 200 on average. This sample collection process persists until convergence is attained.

\subsection{Direct Transmission Neural Network}
To acquire the training data for this network, we compute the probability of transmission for a given incoming direction with the gathered samples:
\begin{equation}
    P(T|\omega_i) = \frac{Count(T|\omega_i)}{Count(All|\omega_i)}.
\end{equation}
We gather samples using the method outlined in \S\ref{ssec:data}, then organize the data into two 22x90 histograms, representing $\omega_i$ with $\theta_i$ and $\phi_i$ bins across a range of 90x360 degrees. The first histogram calculates the number of direct transmission paths, while the second histogram counts the total number of paths. Subsequently, we divide the first histogram by the second to derive the probability map $P(T|\omega_i)$.

Next, we train a lightweight neural network on the probability map. The network, which takes $\omega_i$ as a unit Cartesian vector and predicts $P(T|\omega_i)$, is configured with two hidden layers and follows a 3-7-7-1 structure. The hidden layers utilize the channel-wise Parametric Rectified Linear Unit (PReLU) activation function, and the output layer employs the Sigmoid activation function. The model is trained using the Mean Squared Error (MSE) loss function, coupled with the Adam optimizer. Our network architectures are illustrated in the last column of Fig \ref{fig:pipeline}.

\subsection{Multi-Scattering Neural Network}
\label{sec:multi-scattering}
The neural network is trained on the multi-scattering component of the RDM. To prepare the data for the neural network, it is necessary to isolate the multi-scattering component from the collected samples. Initially, samples with a depth of 0 are removed to exclude the direct transmission samples, along with samples having a depth of 1 $\cap$ $(\omega_o \cdot \textbf{n}) > 0$ to exclude the direct reflection samples. Subsequently, a weighted 4D histogram is computed from the remaining data into 22x90x45x90 bins of $\theta_i$, $\phi_i$, $\theta_o$, and $\phi_o$, each spanning across the respective ranges of 90x360x180x360 degrees. The data is then divided by the number of samples in each incident bin and the solid angle in each outgoing bin to obtain the radiance at each bin. With the multi-scattering component RDM available, samples are randomly drawn from it to generate our training data.

The neural network is configured to accept Cartesian unit vectors $\omega_i$ and $\omega_o$ as inputs and to output r, g, b radiance values. The model incorporates two hidden layers with a 6-21-21-3 structure. The hidden layers utilize channel-wise PReLU activation functions, while the final layer employs the exponential activation function. The model is trained using the Mean Squared Error (MSE) loss function and optimized with the Adam optimizer.

As previously noted in \S\ref{sec:multi-scattering_component}, the multi-scattering component is represented by $S_M(\omega_i, \omega_o)$. However, in practice, the neural network was configured to model the product of $S_M(\omega_i, \omega_o)$ and $(\omega_i \cdot \textbf{n})$. Here, $(\omega_i \cdot \textbf{n})$ represents the cosine foreshortening factor and is inherently included in the RDM as we record the radiance for each $(\omega_i)$ and $(\omega_o)$ directly.

\section{Model Analysis and Ablation}
\label{sec:ablation}
\begin{figure}[b]
    \centering
    \begin{tikzpicture}
    \begin{axis}[
        ybar,
        xlabel={Model Architecture},
        ylabel={Loss for Multi-Scattering Network},
        enlargelimits=0.15,
        legend style={at={(0.5,-0.30)},
        anchor=north,legend columns=-1},
        symbolic x coords={6-21-64-64-21-3, 6-21-64-21-3, 6-21-21-3, 6-16-16-3},
        x tick label style={rotate=20,anchor=east},
        xtick=data,
        % nodes near coords,
        % nodes near coords align={vertical},
        x tick label style={rotate=0},
        ]
        \addplot coordinates {(6-21-64-64-21-3,0.000114421) (6-21-64-21-3,0.000115595) (6-21-21-3,0.000124631) (6-16-16-3,0.000128706)};
        \addplot coordinates {(6-21-64-64-21-3,0.000112205) (6-21-64-21-3,0.000114569) (6-21-21-3,0.000123347) (6-16-16-3,0.000128535)};
        \addplot coordinates {(6-21-64-64-21-3,0.000108866) (6-21-64-21-3,0.000111899) (6-21-21-3,0.000119518) (6-16-16-3,0.000124277)};
        \legend{ELU,ReLU,PReLU}
    \end{axis}
    \end{tikzpicture}
\caption{The performance of various neural network architectures on the multi-scattering component. We demonstrate the effectiveness of channel-wise PReLU compared to common activation functions.}
\label{fig:archgraph}
\end{figure}
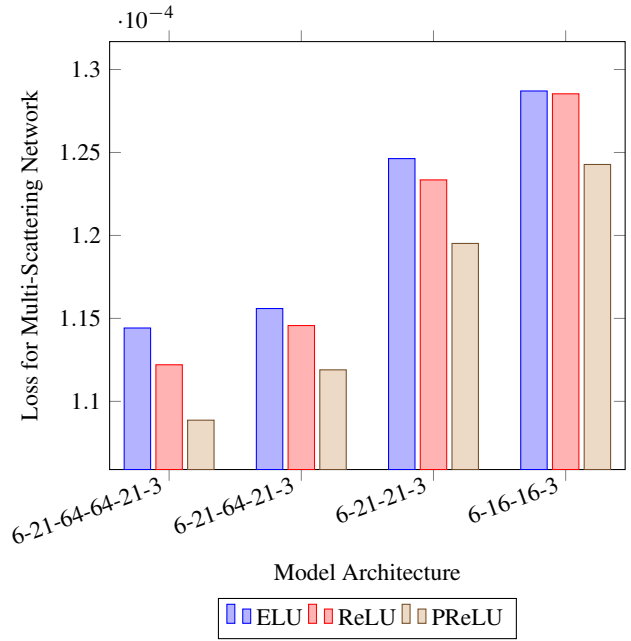

% \begin{figure}
%     \centering
%     \begin{tikzpicture}
%     \begin{axis}[xlabel={Epochs}, ylabel={Loss}, ymode=log]
%     \addlegendentry{Adagrad}
%     \addplot table [x=Epoch, y=Loss, col sep=comma, mark=none] {lossdata/adagrad.csv};
%     \addlegendentry{Adam}
%     \addplot table [x=Epoch, y=Loss, col sep=comma, mark=none] {lossdata/adam.csv};
%     \addlegendentry{RMSprop}
%     \addplot table [x=Epoch, y=Loss, col sep=comma, mark=none] {lossdata/rmsprop.csv};
%     \addlegendentry{SGD}
%     \addplot table [x=Epoch, y=Loss, col sep=comma, mark=none] {lossdata/sgd.csv};
%     \end{axis}
%     \end{tikzpicture}
%     \caption{/-/-T-O-D-O}
%     \label{fig:enter-label}
% \end{figure}
In this section, we perform an ablation study about the neural network used in the multi-scattering component by comparing the performance of the model with different architectures. The model is trained on polyester until convergence (40 epoch). Fig. \ref{fig:archgraph} shows the final loss of various model architectures along with different activation functions. It can be seen that channel-wise PReLU consistently outperforms other activation functions with the same model architecture. This is due to the additional trainable parameters of channel-wise PReLU, which gives the model more flexibility at a negligible increase in runtime cost. It also can be seen that increasing the model weights from our base model to 6-21-64-64-21-3 increases the model size by 10.6 times while only offering a 9\% decrease in loss. From this, we can see that the model does not need to be overly large, and performs well even with a smaller number of weights.

\section{Results}
\label{sec:results}
In this section, we validate our model and evaluate its performance by comparing renderings with our model to reference images generated by rendering the explicit fiber geometry \cite{khungurn2015matching} as well as the hierarchical yarn-based model \cite{Zhu2023yarn}. For all the materials presented, besides polyester, we have used the fiber shading parameters given in Khungurn et al. \cite{khungurn2015matching} which were computed by fitting the parameters to match real-life photographs. The parameters of polyester are determined ad hoc to demonstrate the flexibility of our framework. We then wrap these fibers into yarns with given fiber geometry parameters $N$, $\rho$, and $\alpha$. A summary of the parameters can be found in Table \ref{tab:material_properties}. All images were rendered with path tracing on Mitsuba 3 \cite{Mitsuba3}, including neural network inference, using an Intel Core i7-10750H 6 Core Processor 2.60GHz machine, while neural network training was done on an NVIDIA GeForce RTX4080 (Mobile). The computation time required to gather the RDM is around a minute on an RTX4080 (Mobile). The average time it takes to train a neural network per material for the direct transmission and multi-scattering components are 30 seconds and 30 minutes respectively.

\subsection{Reference Comparisons}
\begin{figure*}[t]
    \centering
    \hspace{-10pt}
    \setlength{\tabcolsep}{1pt}
    \begin{tabular}{cccc}
        Fleece & Half $\rho$ &
        Double $\alpha$ & Polyester \\
        \includegraphics[width = 0.24\linewidth, trim={3cm 2cm 0.8cm 2cm},clip]{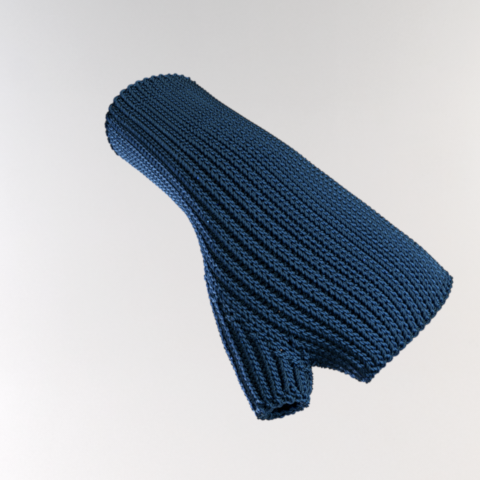} &
        \includegraphics[width = 0.24\linewidth, trim={3cm 2cm 0.8cm 2cm},clip]
        {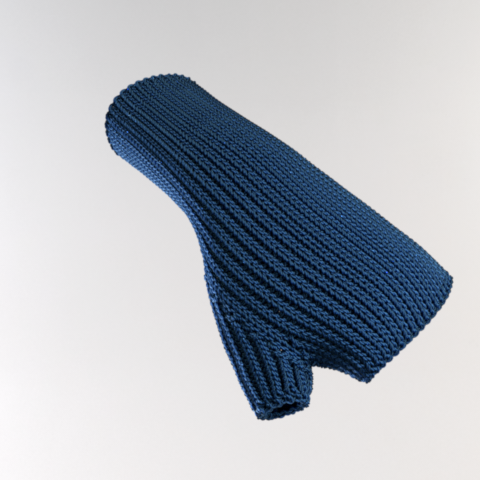} &
        
        \includegraphics[width = 0.24\linewidth, trim={3cm 2cm 0.8cm 2cm},clip] {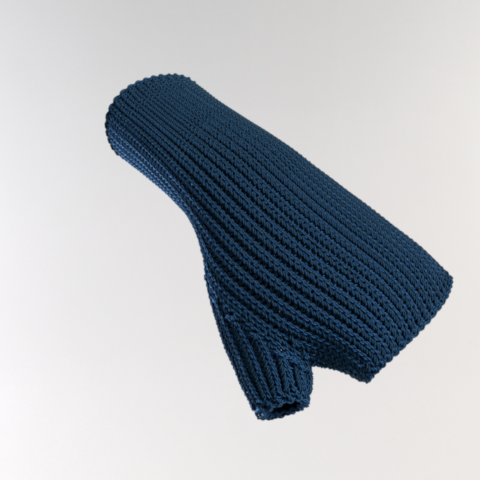} &
        \includegraphics[width = 0.24\linewidth, trim={3cm 2cm 0.8cm 2cm},clip]
        {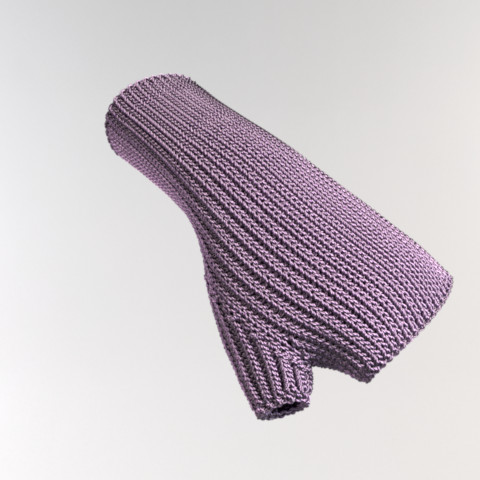} \\
    \end{tabular}
    \caption{Rendering results of a 3-ply glove with fiber parameters; fleece material; fleece with half of the fiber density $\rho$; fleece with double twist factor $\alpha$; polyester material. Please refer to the supplementary video for a comprehensive overview.}
    \label{fig:glovecollage}
\end{figure*}

\emph{3-Ply Knitted Glove} - In this section, we rendered a scene with a 3-ply knitted glove. The base yarn curves defining the glove were taken from \cite{Yukselyarns, Yuksel2012, Wu2017realtime} and wrapped with 3-plies. The plies are then wrapped with fibers procedurally to generate the ground truth image \cite{zhao2016fitting}. For Fig. \ref{fig:teaser} we rendered the scene at a resolution of 1080x1080. Our model matches the ground truth very well and performs 23 times faster while using around 300 times less memory. The scene is lit with an environmental map along with two spherical lights on the top-right and bottom-left corners.  We also rendered the scene with different fiber parameters in Fig. \ref{fig:glovecollage} to highlight the flexibility of our model.

\begin{figure*}
    \centering
    \setlength{\tabcolsep}{3pt} % Adjust the space between columns
    \begin{tabular}{l c c c}
      & Fleece & Half $\rho$ & Double $\alpha$ \\ [5pt] % Adjust the space between rows
      \rotatebox[]{90}{Ours} & 
      \includegraphics[width=.3\linewidth, trim = 0 6cm 0 6cm, clip,  valign=m]{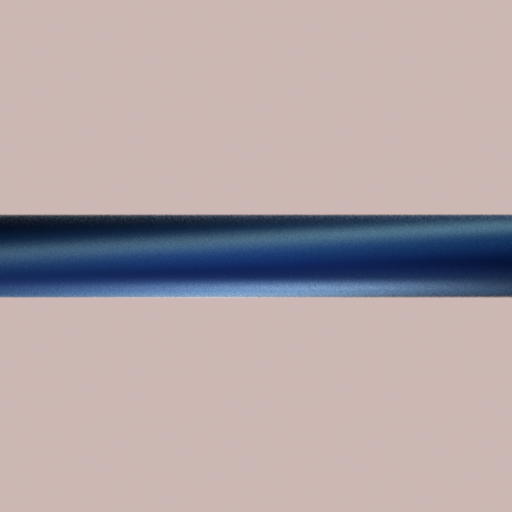} & 
      \includegraphics[width=.3\linewidth, trim = 0 6cm 0 6cm, clip,  valign=m]{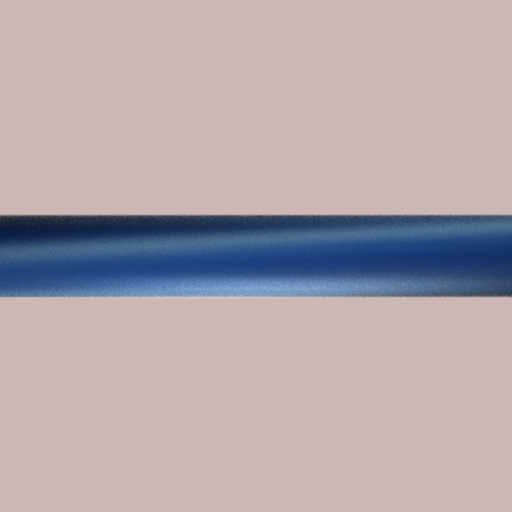} &
      \includegraphics[width=.3\linewidth, trim = 0 6cm 0 6cm, clip,  valign=m]{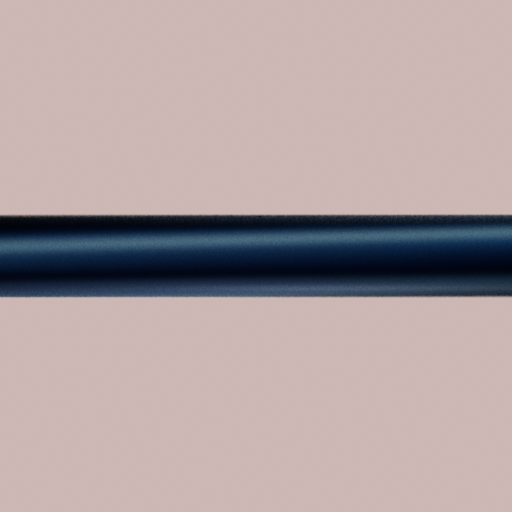}\\ [25pt]
      \rotatebox[]{90}{Ref} & 
      \includegraphics[width=.3\linewidth, trim = 0 6cm 0 6cm, clip,  valign=m]{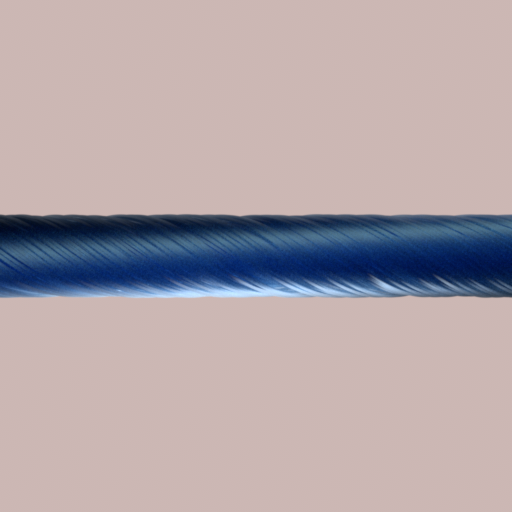} & 
      \includegraphics[width=.3\linewidth, trim = 0 6cm 0 6cm, clip,  valign=m]{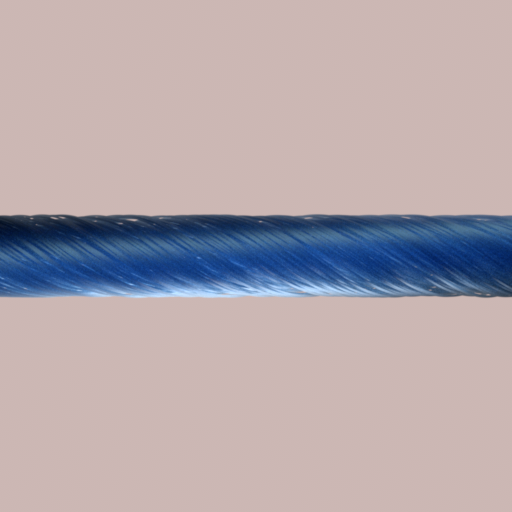} &
      \includegraphics[width=.3\linewidth, trim = 0 6cm 0 6cm, clip,  valign=m]{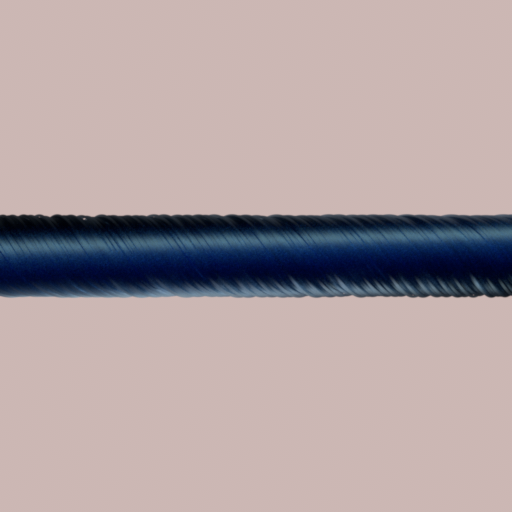}\\ [25pt]
      \rotatebox[]{90}{Ours} & 
      \includegraphics[width=.3\linewidth, trim = 0 6cm 0 6cm, clip,  valign=m]{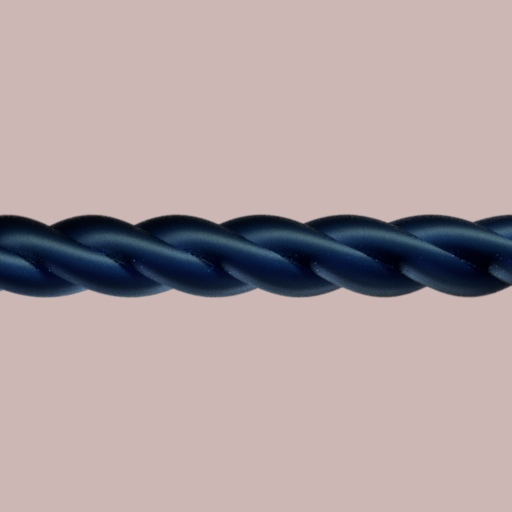} & 
      \includegraphics[width=.3\linewidth, trim = 0 6cm 0 6cm, clip,  valign=m]{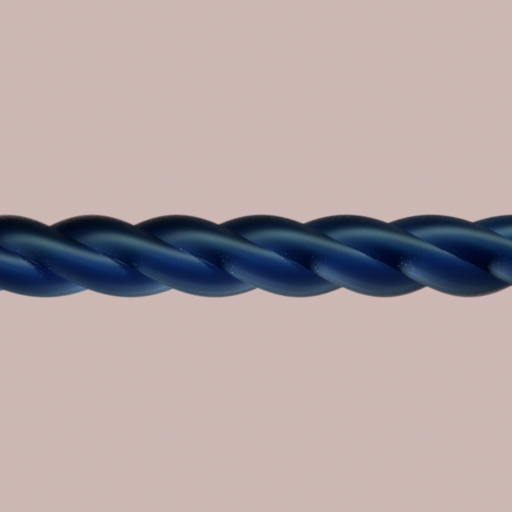} &
      \includegraphics[width=.3\linewidth, trim = 0 6cm 0 6cm, clip,  valign=m]{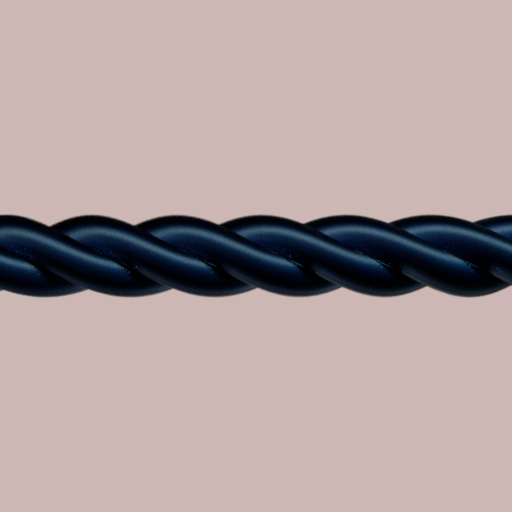}\\ [25pt]
      \rotatebox[]{90}{Ref} & 
      \includegraphics[width=.3\linewidth, trim = 0 6cm 0 6cm, clip,  valign=m]{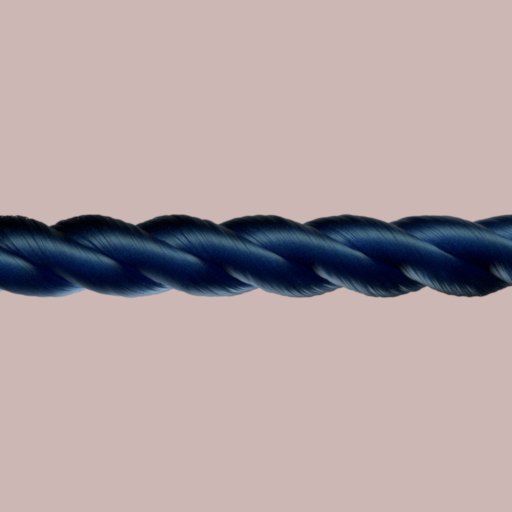} & 
      \includegraphics[width=.3\linewidth, trim = 0 6cm 0 6cm, clip,  valign=m]{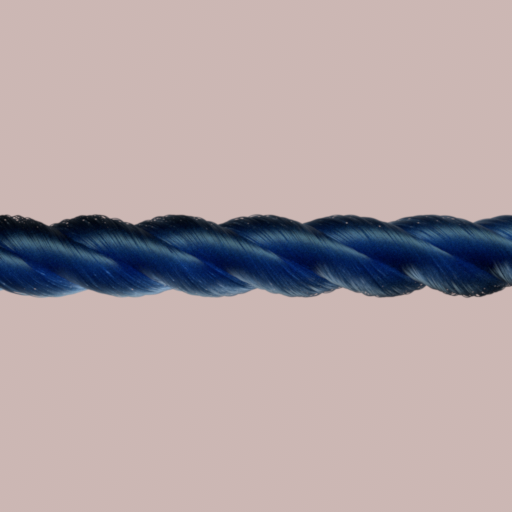} &
      \includegraphics[width=.3\linewidth, trim = 0 6cm 0 6cm, clip,  valign=m]{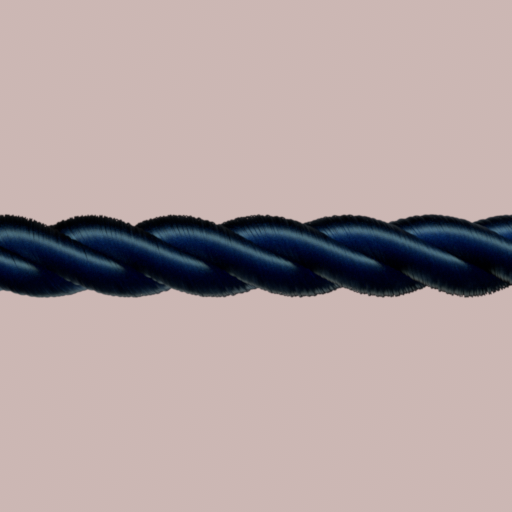}\\ [25pt]
    \end{tabular}
    \caption{Single-yarn comparisons between our aggregated model and the fiber-based ground truth with varying fiber parameters. In this figure, we compare against both 1-ply and 3-ply yarn and demonstrate that our model can accurately recreate the ply level highlights.}
    \label{fig:yarnparams}
\end{figure*}

\emph{Close-up Yarn} - In Fig. \ref{fig:yarnparams}, we compare our model against the reference on a close-up view of a yarn with varying fiber parameters. The scene is rendered at a resolution of 512x512 and then cropped to an appropriate size. The reference is rendered at 1024 spp, while our model is rendered at 64spp and performs on average 30 times faster. Our model can match the overall yarn appearance despite not having explicit fiber geometry. Please note for a multi-ply yarn, our model aggregates the fiber bundle of a single ply and we rely on the renderer to take the ply-ply interactions into account. 

\begin{figure*}
    \centering
    \setlength{\tabcolsep}{3pt} % Adjust the space between columns
    \begin{tabular}{l c c c c l}
      & Reference (ET) & Ours  & Reference (EQ) & SSIM (Ours \& Ref. EQ) & \\ [5pt] % Adjust the space between rows
      \adjustbox{rotate=90,valign=m}{fleece} & 
      \includegraphics[width=.20\linewidth,  valign=m]{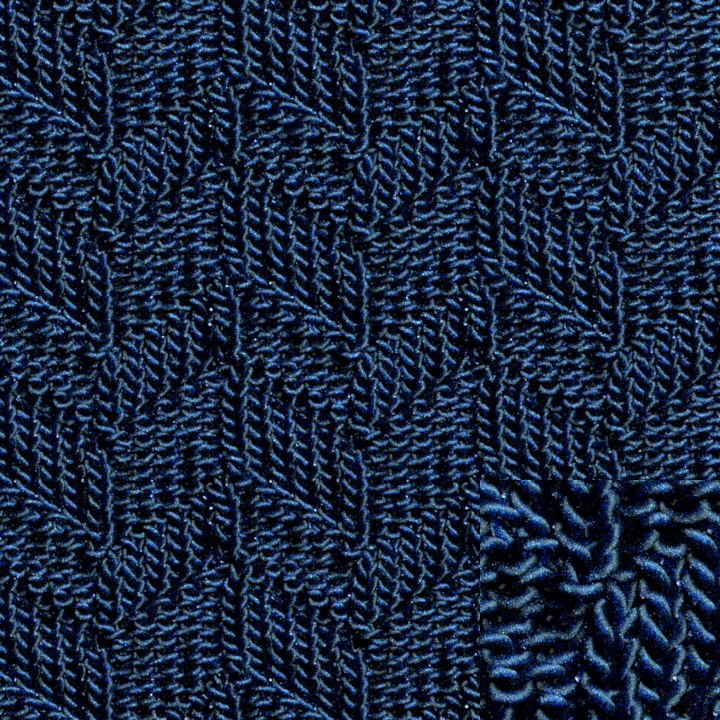} & 
      \includegraphics[width=.20\linewidth,  valign=m]{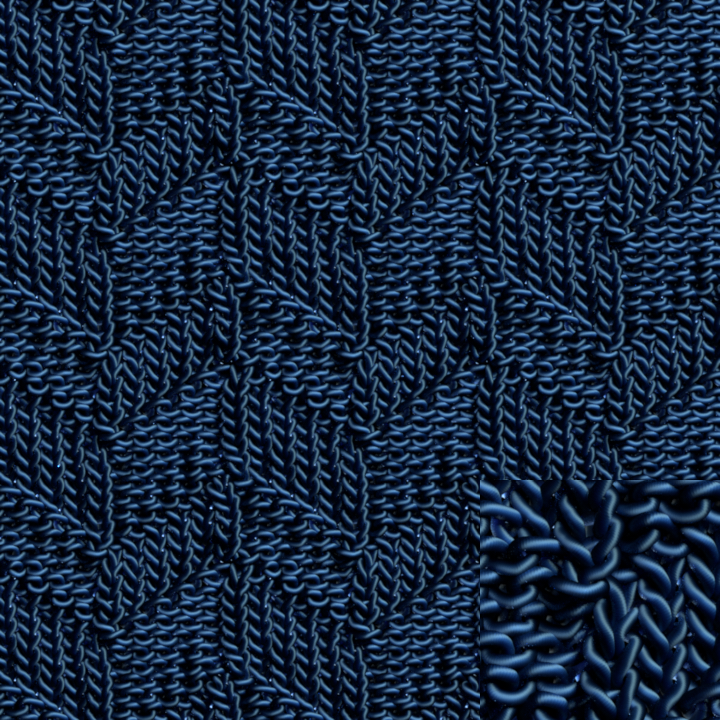} &
      \includegraphics[width=.20\linewidth,  valign=m]{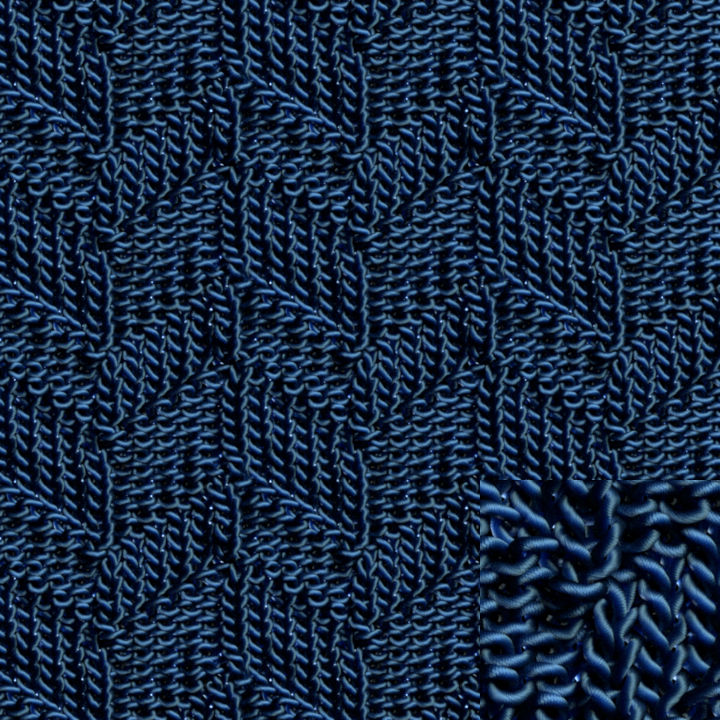} & 
      \includegraphics[width=.20\linewidth,  valign=m]{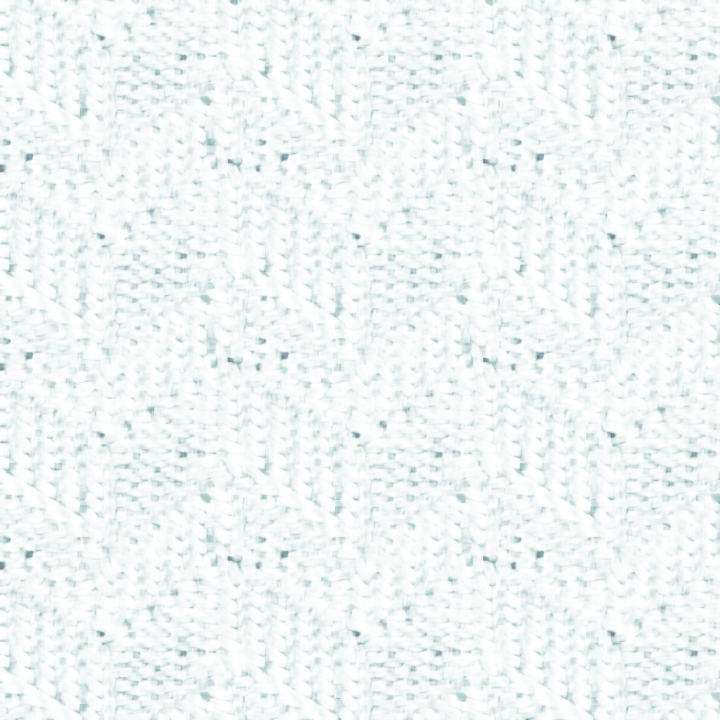} & 
      \includegraphics[height=.20\linewidth,  valign=m]{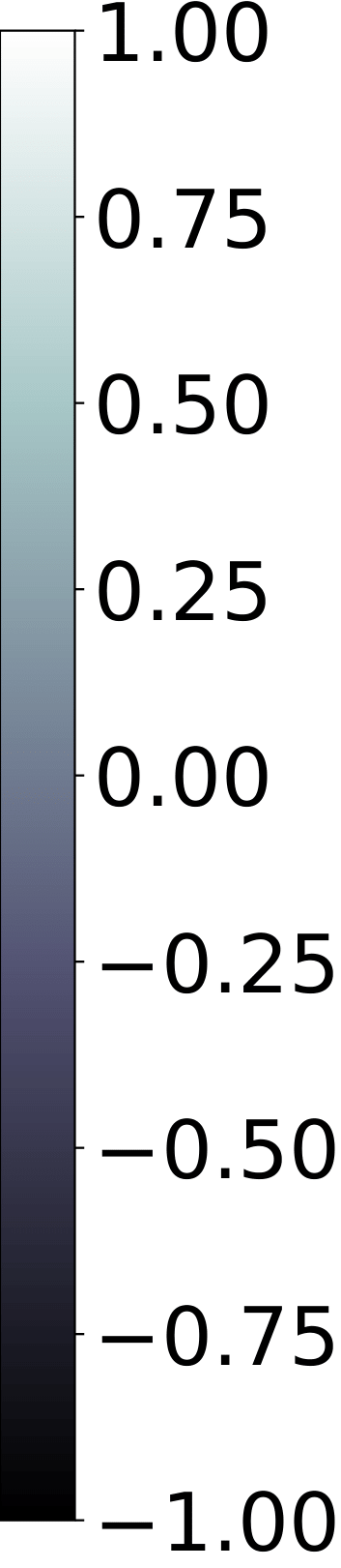}\\ [50pt]
      \adjustbox{rotate=90,valign=m}{silk} & 
      \includegraphics[width=.20\linewidth,  valign=m]{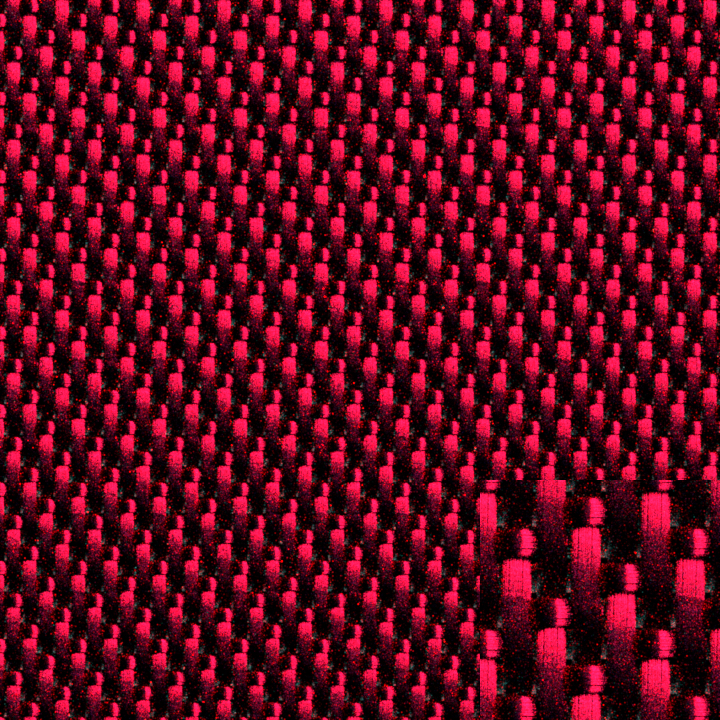} & 
      \includegraphics[width=.20\linewidth,  valign=m]{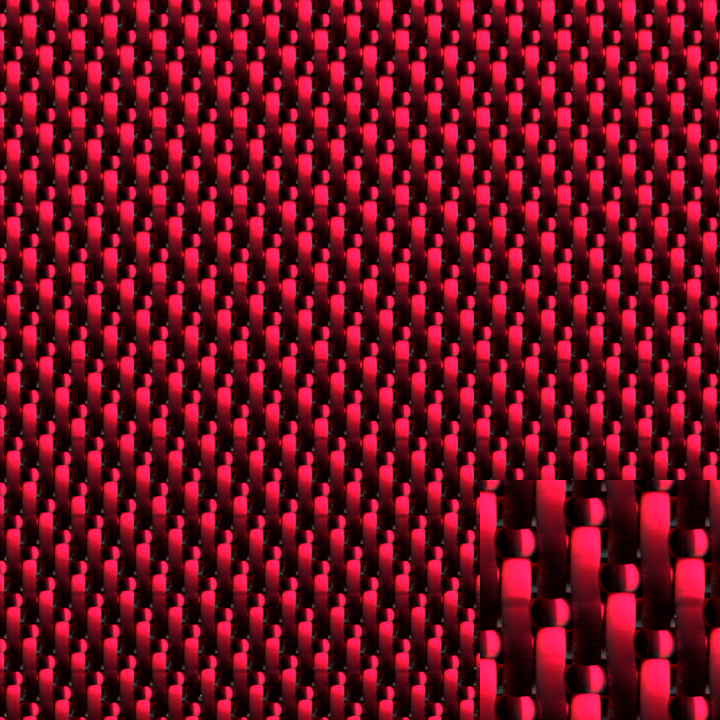} &
      \includegraphics[width=.20\linewidth,  valign=m]{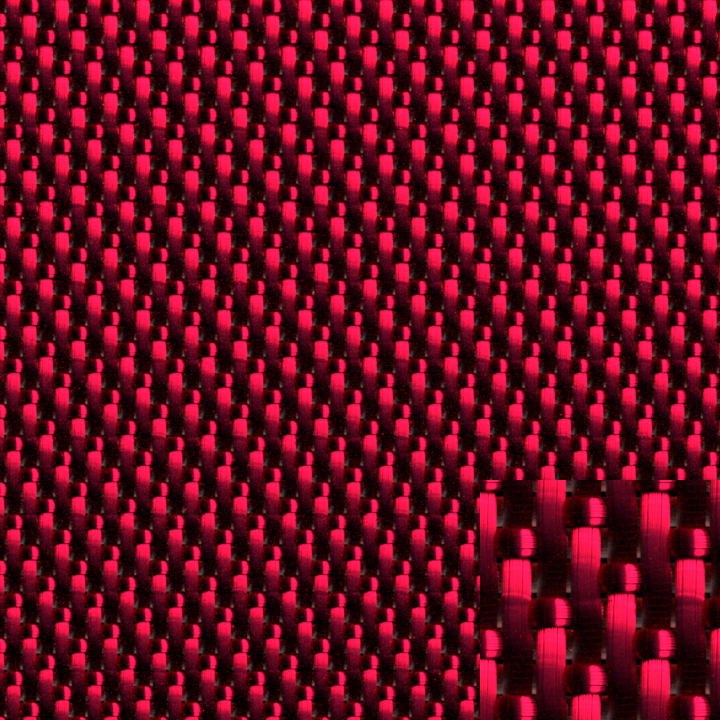} & 
      \includegraphics[width=.20\linewidth,  valign=m]{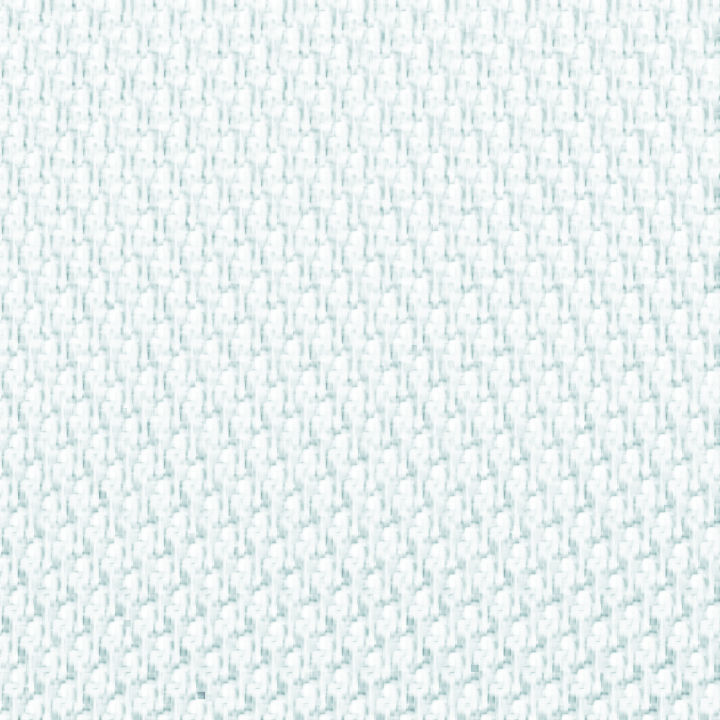} & 
      \includegraphics[height=.20\linewidth,  valign=m]{images/cloth/ssim/colorbar.png}\\ [50pt]
      \adjustbox{rotate=90,valign=m}{polyester} & 
      \includegraphics[width=.20\linewidth,  valign=m]{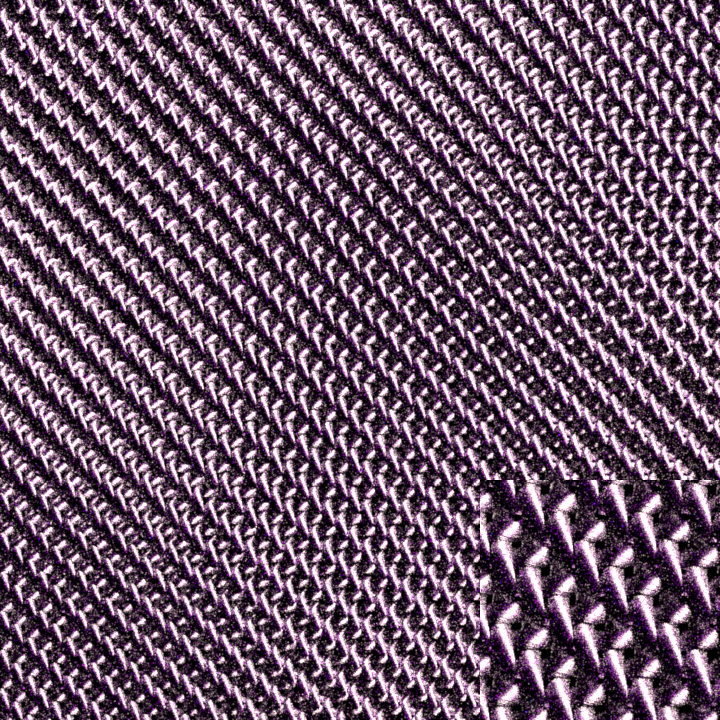} & 
      \includegraphics[width=.20\linewidth,  valign=m]{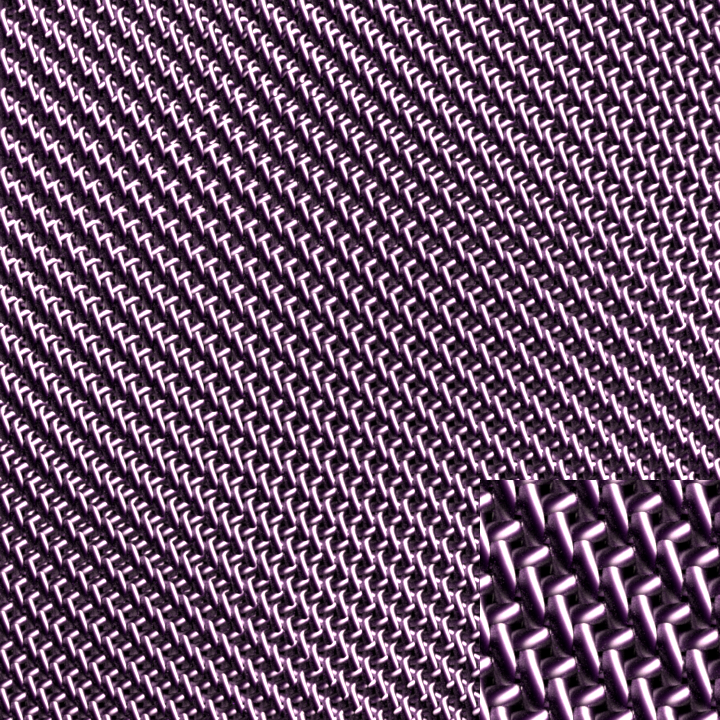} &
      \includegraphics[width=.20\linewidth,  valign=m]{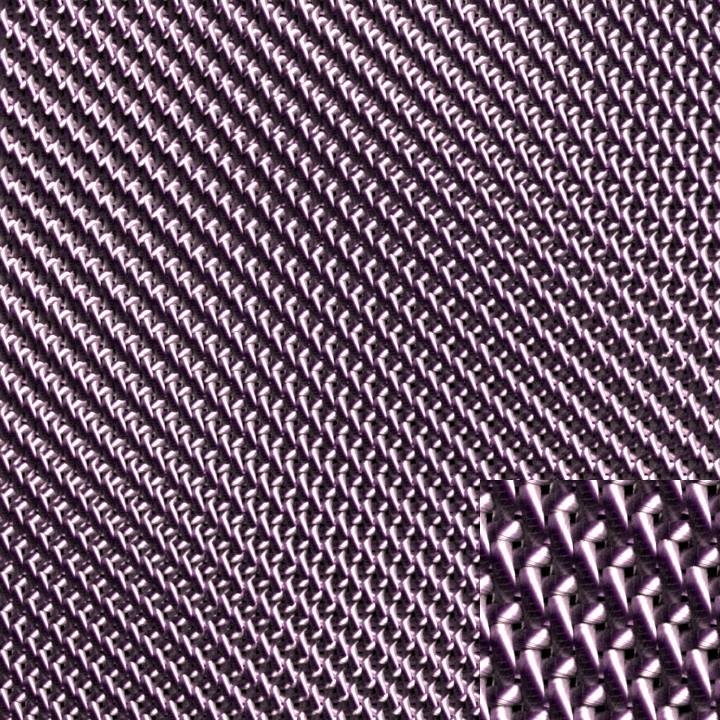} &
      \includegraphics[width=.20\linewidth,  valign=m]{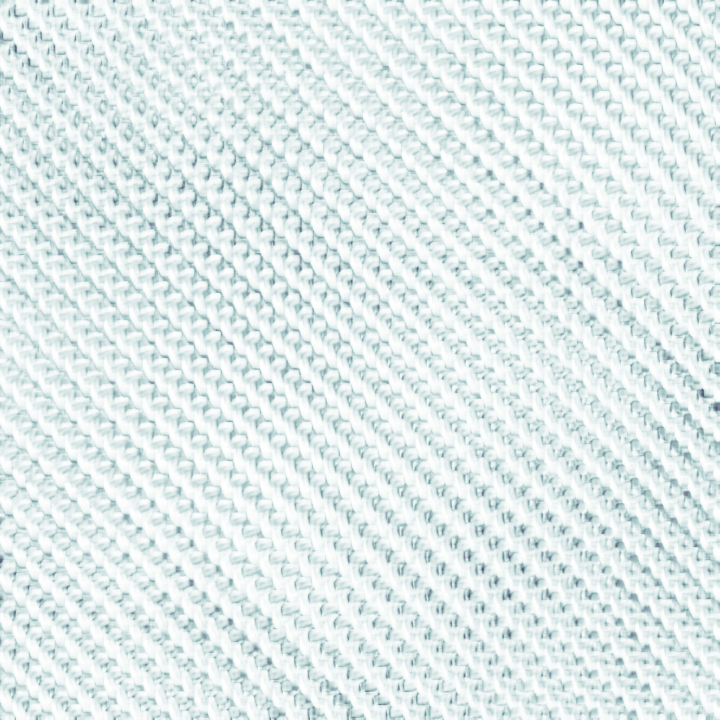} & 
      \includegraphics[height=.20\linewidth,  valign=m]{images/cloth/ssim/colorbar.png}\\ [50pt]
      \adjustbox{rotate=90,valign=m}{cotton} & 
      \includegraphics[width=.20\linewidth,  valign=m]{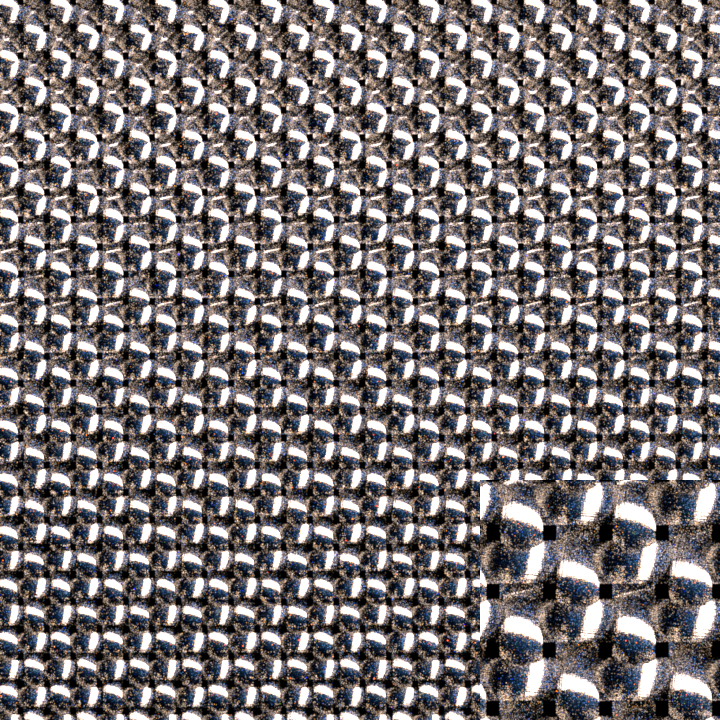} & 
      \includegraphics[width=.20\linewidth,  valign=m]{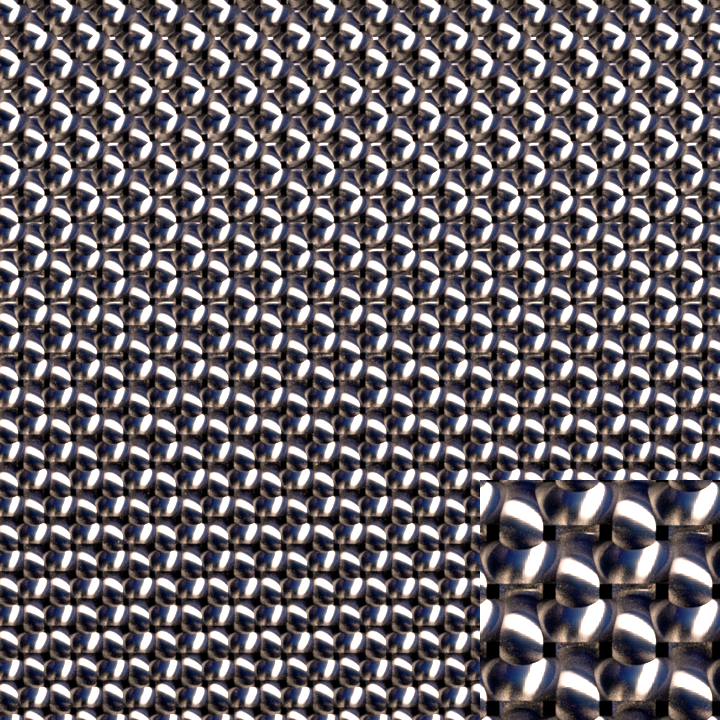} &
      \includegraphics[width=.20\linewidth,  valign=m]{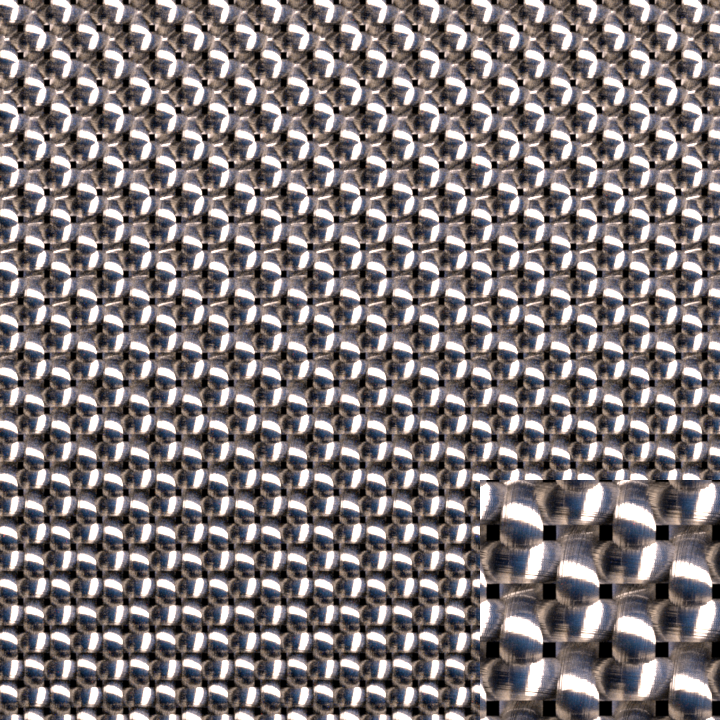} & 
      \includegraphics[width=.20\linewidth,  valign=m]{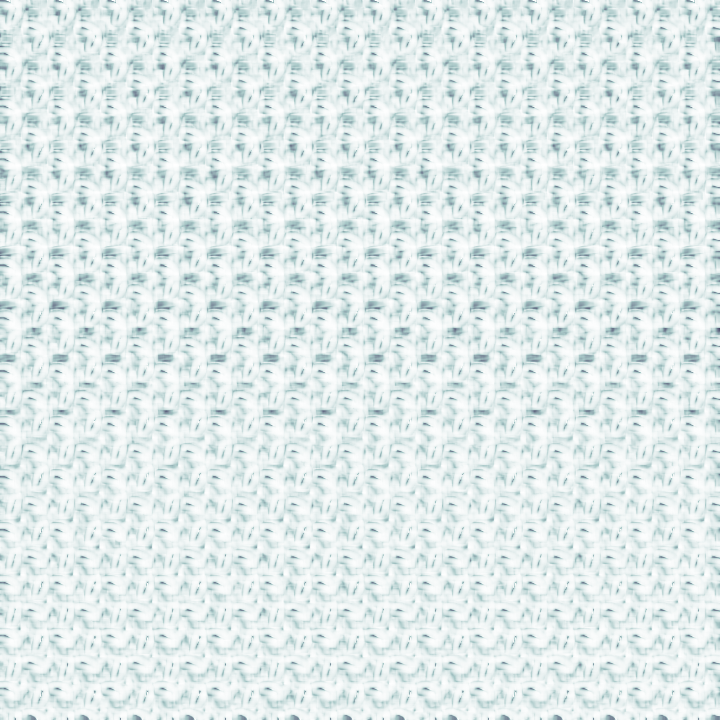} & 
      \includegraphics[height=.20\linewidth,  valign=m]{images/cloth/ssim/colorbar.png}\\ [50pt]
      \adjustbox{rotate=90,valign=m}{gabardine} & 
      \includegraphics[width=.20\linewidth,  valign=m]{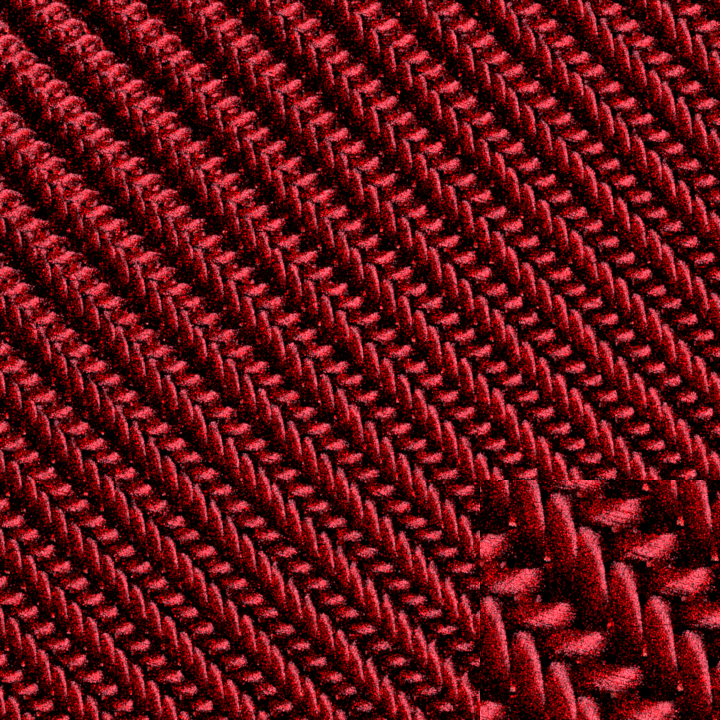} & 
      \includegraphics[width=.20\linewidth,  valign=m]{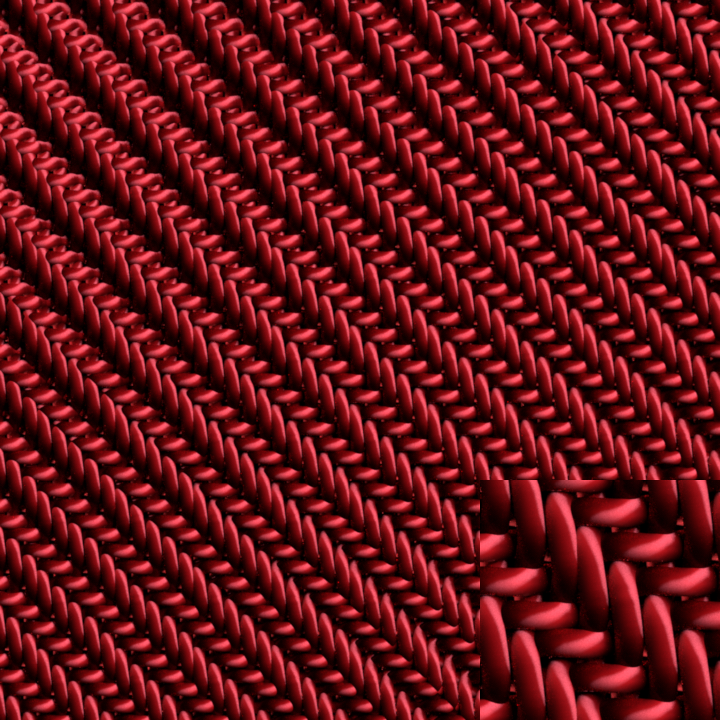} &
      \includegraphics[width=.20\linewidth,  valign=m]{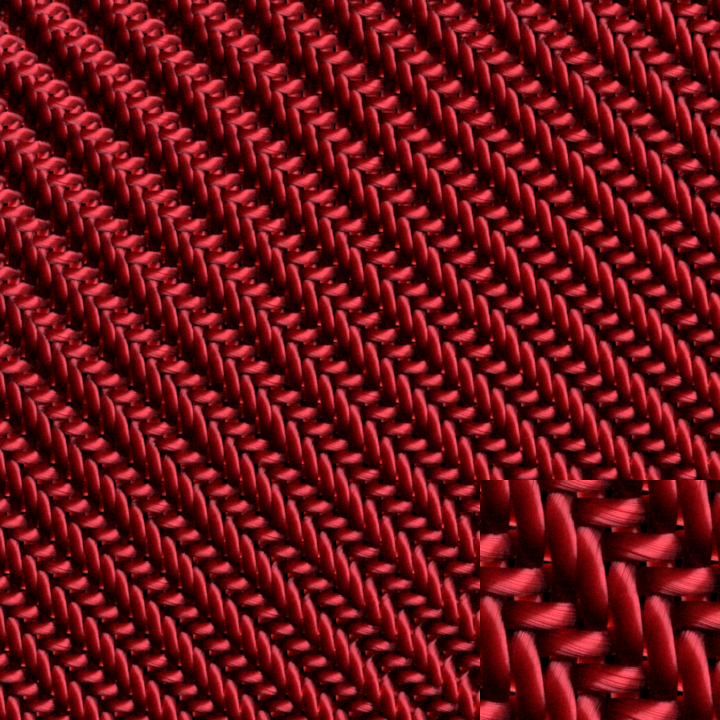} &
      \includegraphics[width=.20\linewidth,  valign=m]{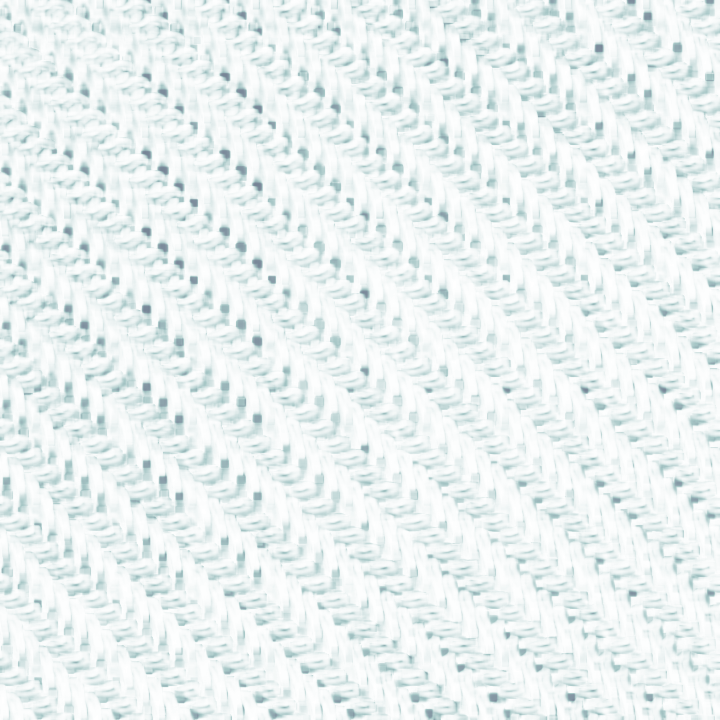} & 
      \includegraphics[height=.20\linewidth,  valign=m]{images/cloth/ssim/colorbar.png}\\ [50pt]
    \end{tabular}
    \caption{The SSIM comparison of our aggregated yarn shading model vs explicit fiber-based models \cite{khungurn2015matching} on knitted and woven fabrics with equal quality. The EQ and ET references are highlighted to showcase their slow and noisy performance, respectively. Our model can accurately replicate the appearance of fiber-based models at a fraction of the time and memory costs.}
    \label{fig:flatcloth}
\end{figure*}

\emph{Woven and Knitted Fabric} - In Fig. \ref{fig:flatcloth}, we rendered our images using the dataset of yarn curves by Leaf et al. \cite{leaf2018stanfordyarn}. The curves were interpolated and tiled into an appropriate size. All the images were rendered at a resolution of 720x720. All the reference images were rendered at 1024spp except for silk and cotton which were rendered at 4096spp as they take longer to converge due to their very high albedo. From our comparisons, our model matches the reference images very well and can accurately recreate yarn-level details even in the absence of explicit fiber geometry. However, although still visually accurate, we do note that cotton has difficulty matching the reference which is discussed further in the limitations section in \S\ref{sec:conclusion}. Our model performs around 11-17 times faster while using around 200-600 times less memory. Please refer to Table \ref{tab:material_comparison} for the full statistics.

\subsection{Comparisons with Zhu et al. 2023}
\begin{figure*}
    \centering
    \setlength{\tabcolsep}{3pt} % Adjust the space between columns
    \begin{tabular}{l c c c c c}
      & fleece & silk  & polyester & cotton & gabardine \\ [5pt] % Adjust the space between rows
      \adjustbox{rotate=90,valign=m}{Reference} & 
      \includegraphics[width=.17\linewidth,  valign=m]{images/cloth/collage/fleece_eq_collage.png} & 
      \includegraphics[width=.17\linewidth,  valign=m]{images/cloth/collage/silk_eq_collage.png} &
      \includegraphics[width=.17\linewidth,  valign=m]{images/cloth/collage/polyester_eq_collage.png} & 
      \includegraphics[width=.17\linewidth,  valign=m]{images/cloth/collage/cotton_eq_collage.png} & 
      \includegraphics[width=.17\linewidth,  valign=m]{images/cloth/collage/gabardine_eq_collage.png}\\ [40pt]
      \adjustbox{rotate=90,valign=m}{Ours} & 
      \includegraphics[width=.17\linewidth,  valign=m]{images/cloth/collage/n-fleece_collage.png} & 
      \includegraphics[width=.17\linewidth,  valign=m]{images/cloth/collage/n-silk_collage.png} &
      \includegraphics[width=.17\linewidth,  valign=m]{images/cloth/collage/n-polyester_collage.png} & 
      \includegraphics[width=.17\linewidth,  valign=m]{images/cloth/collage/n-cotton_collage.png} & 
      \includegraphics[width=.17\linewidth,  valign=m]{images/cloth/collage/n-gabardine_collage.png}\\ [40pt]
      \adjustbox{rotate=90,valign=m}{\cite{Zhu2023yarn}} & 
      \includegraphics[width=.17\linewidth,  valign=m]{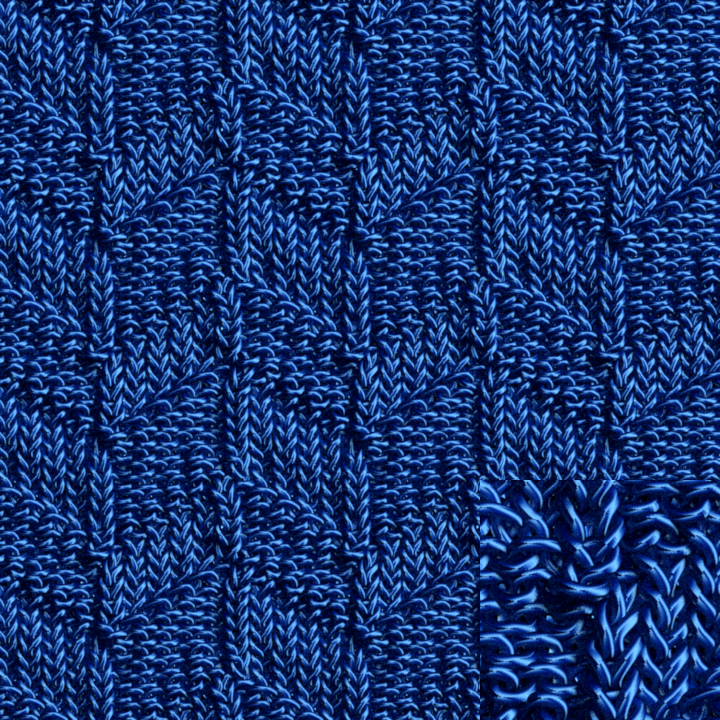} & 
      \includegraphics[width=.17\linewidth,  valign=m]{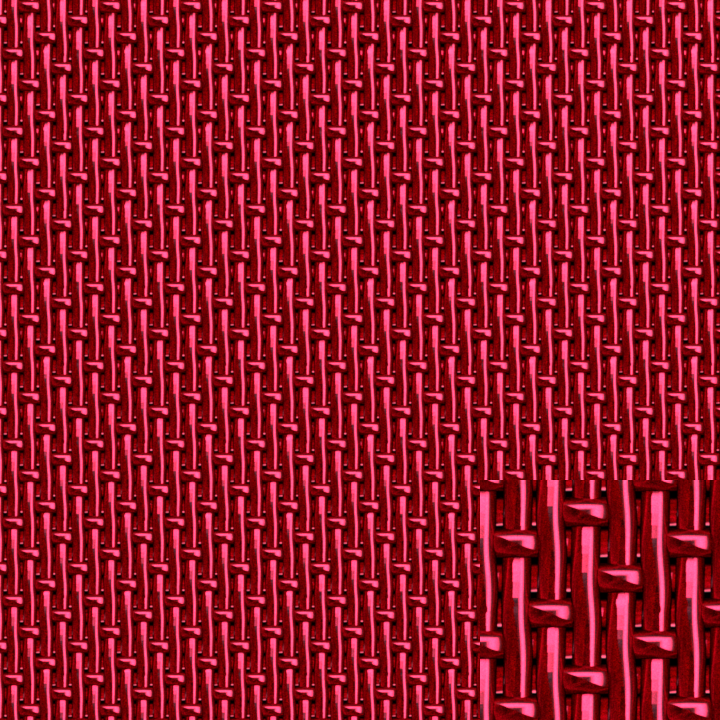} &
      \includegraphics[width=.17\linewidth,  valign=m]{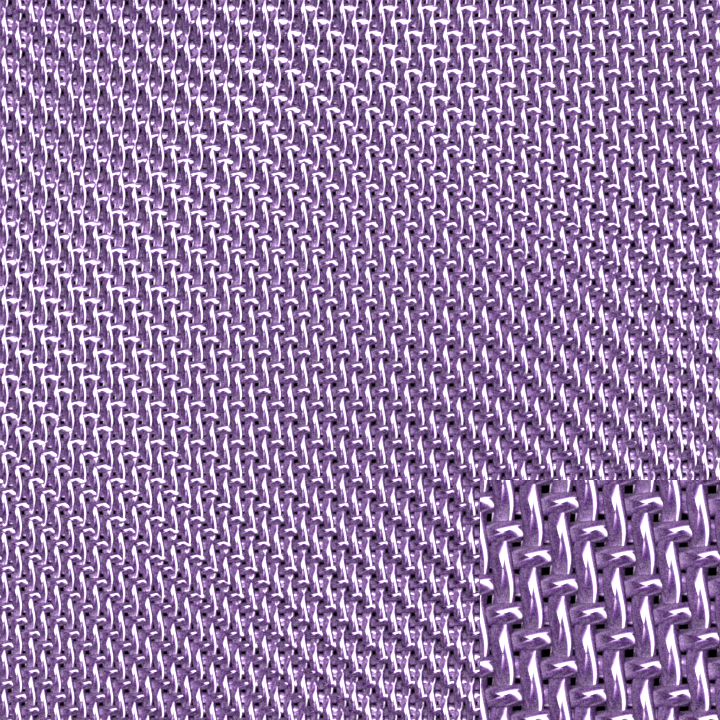} & 
      \includegraphics[width=.17\linewidth,  valign=m]{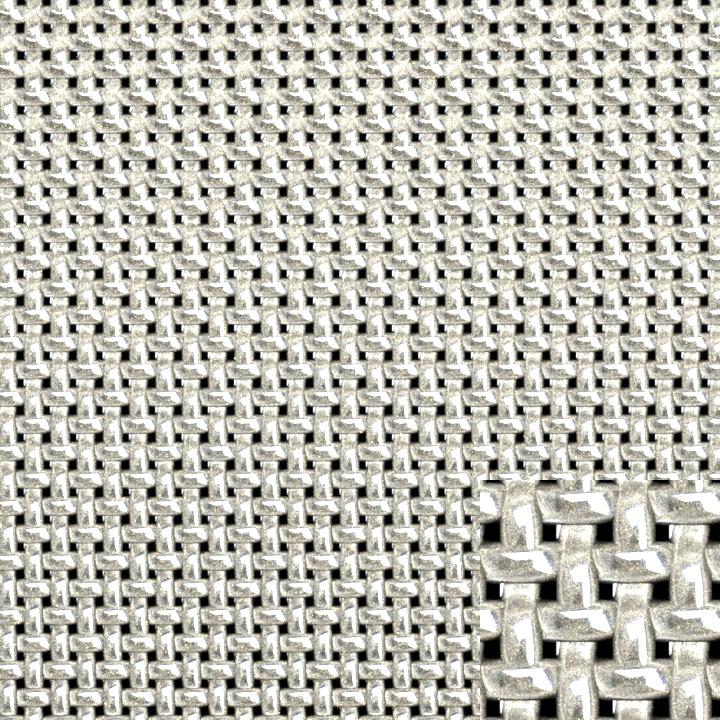} & 
      \includegraphics[width=.17\linewidth,  valign=m]{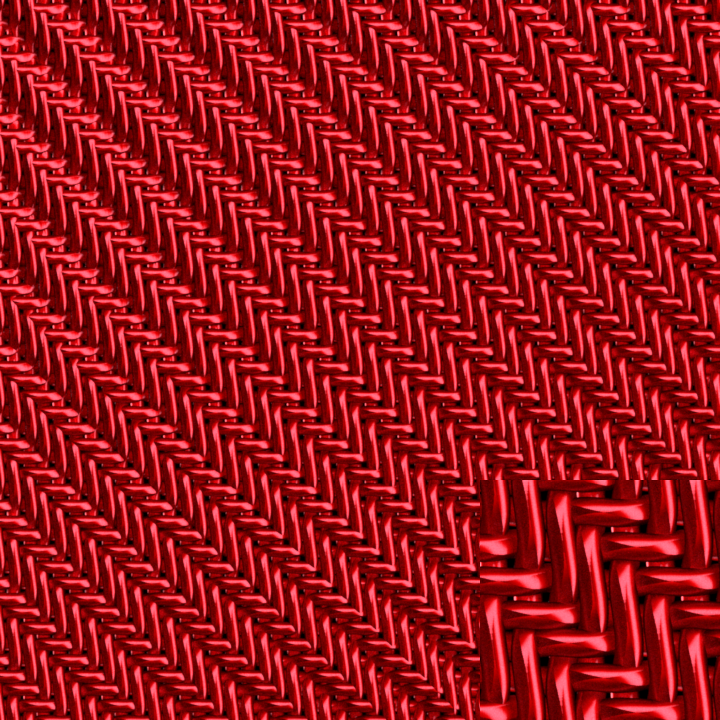}\\ [40pt]
    \end{tabular}
    \caption{A comparison of ours with state-of-the-art model \cite{Zhu2023yarn} in equal quality, using reference's parameters for aggregation}
    \label{fig:zhucomparison}
\end{figure*}

As depicted in Fig. \ref{fig:zhucomparison}, we demonstrate that our approach not only achieves faster rendering speeds, as detailed in Table \ref{tab:material_comparison}, but also more accurately replicates the reference fiber-based appearance model by Khungurn et al. \cite{khungurn2015matching}. Our model's superiority is due to our neural data-driven methodology that adapts more flexibly, allowing for an exact fit to the reference. In contrast, \cite{Zhu2023yarn} uses an approximated fiber appearance model, which does not model Fresnel effects, and often requiring manual adjustments to align with the reference model. Notably, we use the exact same set of parameters and values across the three models (reference, ours, and \cite{Zhu2023yarn}) without any post-tweaking. 

% \emph{Comparisons with \cite{Zhu2023yarn}} - From our experiments in Fig. \ref{fig:flatcloth}, we demonstrate that our approach outperforms \cite{Zhu2023yarn} in terms of rendering speed as detailed in Table \ref{tab:material_comparison}. Additionally, our model is able to better match the reference fiber-based appearance model by Khungurn et al. \cite{khungurn2015matching}, as we utilize a data-driven approach that allows for better flexibility by overfitting to the reference. In contrast, \cite{Zhu2023yarn} is limited to the fiber appearance model described in their paper, and in practice, requires some manual tweaking to match the reference, as using the parameters directly would yield suboptimal results. 

% \begin{table*}[h]
%     \centering
%     \begin{tabular}{lrrrrrrrrr}
%         \toprule
%         Material & \multicolumn{5}{c}{Time (s)} & \multicolumn{4}{c}{Memory (MB)} \\
%         \cmidrule(lr){2-6} \cmidrule(lr){7-10}
%         & Ref & \cite{Zhu2023yarn} & Ratio & Ours & Ratio  & Ref & \cite{Zhu2023yarn} & Ours & Ratio \\
%         \midrule
%         fleece    & 12631  & 3263 & $\times3.9$ & \textbf{952}  & $\mathbf{\times13.3}$ & 6032 & 20 & \textbf{20} & $\mathbf{\times307}$ \\
%         silk      & 31223 & 20831 & $\times1.5$ & \textbf{2908} & $\mathbf{\times10.7}$ & 6216 & 20 & \textbf{20} & $\mathbf{\times307}$ \\
%         polyester & 8229  & 11932 & $\times0.7$ & \textbf{612}  & $\mathbf{\times13.5}$ & 5935 & 29 & \textbf{29} & $\mathbf{\times202}$ \\
%         cotton    & 44836 & 24167 & $\times1.9$ & \textbf{2626} & $\mathbf{\times17.1}$ & 4391 & 7 & \textbf{7}  & $\mathbf{\times607}$ \\
%         gabardine & 11427 & 13024 & $\times0.9$ & \textbf{785}  & $\mathbf{\times14.6}$ & 9340 & 20 & \textbf{20} & $\mathbf{\times461}$ \\
%         \bottomrule
%     \end{tabular}
%     \caption{Performance Statistics for Fig. \ref{fig:flatcloth}. All rendering times were counted at equal quality (EQ).}
%     \label{tab:material_comparison}
% \end{table*}

\begin{table}[t]
    \centering
    \small
    \caption{Performance Statistics for Fig. \ref{fig:flatcloth}. All rendering times were counted at equal quality (EQ).}
    \label{tab:material_comparison}
    \hspace{-0.7cm}
    \begin{tabular}{lrrrrrr}
        \toprule
        Material & \multicolumn{3}{c}{Time (s)} & \multicolumn{3}{c}{Memory (MB)} \\
        \cmidrule(lr){2-4} \cmidrule(lr){5-7}
        & Ref & \cite{Zhu2023yarn} & Ours  & Ref & \cite{Zhu2023yarn} & Ours  \\
        \midrule
        fleece    & 12631  & 3263 & \textbf{952}  & 6032 & 20 & \textbf{20} \\
        silk      & 31223 & 20831 & \textbf{2908} & 6216 & 20 & \textbf{20} \\
        polyester & 8229  & 11932 & \textbf{612}  & 5935 & 29 & \textbf{29} \\
        cotton    & 44836 & 24167 & \textbf{2626} & 4391 & 7 & \textbf{7}  \\
        gabardine & 11427 & 13024 & \textbf{785}  & 9340 & 20 & \textbf{20}  \\
        \bottomrule
    \end{tabular}
\end{table}

\begin{table*}[t]
    \centering
    \caption{Fiber parameters used throughout our paper. The shading parameters are based on matched fibers from \cite{khungurn2015matching}, and the geometrical parameters are set on an ad hoc basis}.
    \label{tab:material_properties}
    \begin{tabular}{lcccccccc}
        \toprule
        & $N$ & $\rho$ & $\alpha$ & $C_R$ & $C_{TT}$ & $\beta_R$ & $\beta_{TT}$ & $\gamma_{TT}$ \\
        \midrule
        fleece    & 300 & 0.30 & 0.24 & 0.040, 0.087, 0.087 & 0.452, 0.725, 0.948 & 7.238  & 10.000 & 25.989 \\
        silk      & 300 & 0.20 & 0.00 & 0.745, 0.008, 0.070 & 0.620, 0.553, 0.562 & 1.000  & 10.000 & 19.823 \\
        polyester & 200 & 0.40 & 0.20 & 0.700, 0.700, 0.700 & 0.600, 0.000, 0.800 & 5.238  & 20.000 & 25.000 \\
        cotton    & 600 & 0.35 & 0.06 & 0.989, 0.959, 0.874 & 0.999, 0.999, 0.999 & 1.000  & 27.197 & 38.269 \\
        gabardine & 450 & 0.25 & 0.12 & 0.185, 0.047, 0.069 & 0.999, 0.330, 0.354 & 2.141  & 10.000 & 23.548 \\
        \bottomrule
    \end{tabular}
\end{table*}
% \begin{table}[h!]
%     \centering
%     \small
%     \begin{tabular}{lccccccc}
%         \toprule
%         type, $N$ & $\rho$ & $\alpha$ & $C_R$ & $C_{TT}$ & $\beta_R$ & $\beta_{TT}$ & $\gamma_{TT}$ \\
%         \midrule
%         fleece    , 300 & 0.30 & 0.24 & 0.04, 0.08, 0.09 & 0.45, 0.72, 0.95 & 7.24  & 10.00 & 25.99 \\
%         silk      , 300 & 0.20 & 0.00 & 0.74, 0.01, 0.07 & 0.62, 0.55, 0.56 & 1.00  & 10.00 & 19.82 \\
%         polyester , 200 & 0.40 & 0.20 & 0.70, 0.70, 0.70 & 0.60, 0.00, 0.80 & 5.24  & 20.00 & 25.00 \\
%         cotton    , 600 & 0.35 & 0.06 & 0.99, 0.96, 0.87 & 0.99, 0.99, 0.99 & 1.00  & 27.19 & 38.27 \\
%         gabardine , 450 & 0.25 & 0.12 & 0.18, 0.05, 0.07 & 0.99, 0.33, 0.35 & 2.14  & 10.00 & 23.55 \\
%         \bottomrule
%     \end{tabular}
%     \caption{Fiber parameters used throughout our paper. The shading parameters are based on matched fibers from \cite{khungurn2015matching}, and the geometrical parameters are set on an ad hoc basis}.
%     \label{tab:material_properties}
% \end{table}

\section{Conclusion and Discussion}
\label{sec:conclusion}
\emph{Limitations} - Our final aggregated ply shading model assumes that the light scattering enters and exits from the same spot and does not exhibit subsurface scattering. Based on our experiments, this is true for most fabrics except for fibers with very high albedo (such as cotton with 0.999) as they exhibit significantly more bounces per sample and hence travel more throughout the yarn, causing the exit point to be far from the the enter point. While this assumption satisfies most of our cloth types, we left a more accurate distribution of the exit point as a future study. Furthermore, our model assumes the appearance of the yarn is not spatially varying, and is unable to handle spatially varying yarn colors such as dyed cloth. Lastly, our model requires re-training to alter the yarn parameters, which might limit its use in interactive design and modelling for artists.

\emph{Future Works} - Besides addressing the limitations above, a straightforward extension can include the training and fitting of more complex fiber distributions and scattering as the neural network has the potential to learn any complex distributions. Additionally, we would like to develop and leverage an auto-encoder architecture similar to \cite{sztrajman2021nbrdf} to instantly interpolate our fitted yarn models with different fiber parameters to provide additional flexibility to designers and artists. Although our model performs well with an analytically fitted importance sampling lobe, we are interested in seeing if neural importance sampling methods could be used to further improve convergence \cite{xu2023neusample}. We also would like to extend our method to support efficient level-of-detail simplification. This involves simplifying our model into a 3-dimensional BCSDF using a smaller neural network for far-field views, specifically when the width of the yarn is less than a pixel.

\emph{Conclusions} - In this paper, we presented a novel aggregated shading framework by leveraging the flexibility and generality of neural networks to model the light interactions with a bundle of fibers i.e.\ ply. Our model can replicate the appearance of many fabrics while running significantly faster and requiring less memory. Through observations of the RDM, we also derived an analytical approximation and importance sampling scheme to further improve the rate of convergence of our model. Finally, our fitted model can be applied to any yarn geometry instantly, providing greater flexibility in designing fabrics.

%-------------------------------------------------------------------------
% % bibtex
% \bibliographystyle{eg-alpha-doi}  
% \bibliography{main}        

% biblatex with biber
\printbibliography  

@String{tog = "ACM TOG"}

@misc{Yukselyarns,
  title={Yarn-level Cloth Models},
  author={Cem Yuksel},
  year={2020},
  howpublished={\url{http://www.cemyuksel.com/research/yarnmodels}},
}

@article{Yuksel2012,
   author       = {Cem Yuksel and Jonathan M. Kaldor and Doug L. James and Steve Marschner},
   title        = {Stitch Meshes for Modeling Knitted Clothing with Yarn-Level Detail},
   journal      = {ACM Transactions on Graphics (Proceedings of SIGGRAPH 2012)},
   year         = {2012},
   volume       = {31},
   number       = {3},
   pages        = {37:1--37:12},
   articleno    = {37},
   numpages     = {12},
   location     = {Los Angeles, USA},
   url          = {http://doi.acm.org/10.1145/2185520.2185533},
   doi          = {10.1145/2185520.2185533},
   publisher    = {ACM Press},
   address      = {New York, NY, USA},
}

@inproceedings{Wu2017realtime, 
author = {Kui Wu and Cem Yuksel}, 
title = {Real-time Fiber-level Cloth Rendering}, 
booktitle = {ACM SIGGRAPH Symposium on Interactive 3D Graphics and Games (I3D 2017)}, 
year = {2017}, 
numpages = {8}, 
location = {San Francisco, CA}, 
isbn = {978-1-4503-4886-7/17/03}, 
url = {http://doi.acm.org/10.1145/3023368.3023372}, 
doi = {10.1145/3023368.3023372}, 
publisher = {ACM}, 
address = {New York, NY, USA}, 
}

@software{Mitsuba3,
    title = {Mitsuba 3 renderer},
    author = {Wenzel Jakob and Sébastien Speierer and Nicolas Roussel and Merlin Nimier-David and Delio Vicini and Tizian Zeltner and Baptiste Nicolet and Miguel Crespo and Vincent Leroy and Ziyi Zhang},
    note = {https://mitsuba-renderer.org},
    version = {3.1.1},
    year = 2022
}

@article{Zhu2022fur,
author = {Zhu, Junqiu and Zhao, Sizhe and Wang, Lu and Xu, Yanning and Yan, Ling-Qi},
title = {Practical Level-of-Detail Aggregation of Fur Appearance},
year = {2022},
issue_date = {July 2022},
publisher = {Association for Computing Machinery},
address = {New York, NY, USA},
volume = {41},
number = {4},
issn = {0730-0301},
url = {https://doi.org/10.1145/3528223.3530105},
doi = {10.1145/3528223.3530105},
journal = {ACM Trans. Graph.},
month = {jul},
articleno = {47},
numpages = {17}
}

@misc{khattar2024multiscale,
      title={A Multi-scale Yarn Appearance Model with Fiber Details}, 
      author={Apoorv Khattar and Junqui Zhu and Emiliano Padovani and Jean-Marie Aurby and Marc Droske and Ling-Qi Yan and Zahra Montazeri},
      year={2024},
      eprint={2401.12724},
      archivePrefix={arXiv},
      primaryClass={cs.GR}
}

@article {Zhu2023yarn,
journal = {Computer Graphics Forum},
title = {{A Practical and Hierarchical Yarn-based Shading Model for Cloth}},
author = {Zhu, Junqiu and Montazeri, Zahra and Aubry, Jean-Marie and Yan, Ling-Qi and Weidlich, Andrea},
year = {2023},
publisher = {The Eurographics Association and John Wiley & Sons Ltd.},
ISSN = {1467-8659},
DOI = {10.1111/cgf.14894}
}

@article{loubet2018,
  TITLE = {{A new microflake model with microscopic self-shadowing for accurate volume downsampling}},
  AUTHOR = {Loubet, Guillaume and Neyret, Fabrice},
  URL = {https://hal.science/hal-01702000},
  JOURNAL = {{Computer Graphics Forum}},
  PUBLISHER = {{Wiley}},
  VOLUME = {37},
  NUMBER = {2},
  PAGES = {111-121},
  YEAR = {2018},
  MONTH = May,
  DOI = {10.1111/cgf.13346},
  HAL_ID = {hal-01702000},
  HAL_VERSION = {v1},
}

@article{Marschner2003,
 author = {Marschner, Stephen R. and Jensen, Henrik Wann and Cammarano, Mike and Worley, Steve and Hanrahan, Pat},
 title = {Light Scattering from Human Hair Fibers},
 year = {2003},
 issue_date = {July 2003},
 publisher = {Association for Computing Machinery},
 address = {New York, NY, USA},
 volume = {22},
 number = {3},
 issn = {0730-0301},
 url = {https://doi.org/10.1145/882262.882345},
 doi = {10.1145/882262.882345},
 journal = {ACM Trans. Graph.},
 month = jul,
 pages = {780–791},
 numpages = {12}
}

@article{Montazeri2021mechanics,
title = "Mechanics-Aware Modeling of Cloth Appearance",
author = "Zahra Montazeri and Chang Xiao and Yun Fei and Changxi Zheng and Shuang Zhao",
year = "2021",
month = jan,
day = "1",
doi = "10.1109/TVCG.2019.2937301",
language = "English",
pages = "137 -- 150",
journal = "IEEE Transactions on Visualization and Computer Graphics",
issn = "1077-2626",
publisher = "IEEE Computer Society ",
}

@inproceedings{adabala2003,
author = {Adabala, Neeharika and Magnenat-Thalmann, Nadia and Fei, Guangzheng},
title = {Real-Time Rendering of Woven Clothes},
year = {2003},
isbn = {1581135696},
publisher = {Association for Computing Machinery},
address = {New York, NY, USA},
url = {https://doi.org/10.1145/1008653.1008663},
doi = {10.1145/1008653.1008663},
pages = {41–47},
numpages = {7},
location = {Osaka, Japan},
series = {VRST '03}
}

@inproceedings {montazeri2021practical,
booktitle = {Eurographics Symposium on Rendering - DL-only Track},
editor = {Bousseau, Adrien and McGuire, Morgan},
title = {{Practical Ply-Based Appearance Modeling for Knitted Fabrics}},
author = {Montazeri, Zahra and Gammelmark, Søren and Jensen, Henrik Wann and Zhao, Shuang},
year = {2021},
publisher = {The Eurographics Association},
ISSN = {1727-3463},
ISBN = {978-3-03868-157-1},
DOI = {10.2312/sr.20211297}
}

@article{montazeri2020practical,
  title={A practical ply-based appearance model of woven fabrics},
  author={Montazeri, Zahra and Gammelmark, S{\o}ren B and Zhao, Shuang and Jensen, Henrik Wann},
  journal={ACM Transactions on Graphics (TOG)},
  volume={39},
  number={6},
  pages={1--13},
  year={2020},
  publisher={ACM New York, NY, USA}
}

@article{zhao2016fitting,
  title={Fitting procedural yarn models for realistic cloth rendering},
  author={Zhao, Shuang and Luan, Fujun and Bala, Kavita},
  journal={ACM Transactions on Graphics (TOG)},
  volume={35},
  number={4},
  pages={1--11},
  year={2016},
  publisher={ACM New York, NY, USA}
}

@article{zhao2011building,
  title={Building volumetric appearance models of fabric using micro CT imaging},
  author={Zhao, Shuang and Jakob, Wenzel and Marschner, Steve and Bala, Kavita},
  journal={ACM Transactions on Graphics (TOG)},
  volume={30},
  number={4},
  pages={1--10},
  year={2011},
  publisher={ACM New York, NY, USA}
}

@article{khungurn2015matching,
  title={Matching Real Fabrics with Micro-Appearance Models.},
  author={Khungurn, Pramook and Schroeder, Daniel and Zhao, Shuang and Bala, Kavita and Marschner, Steve},
  journal={ACM Trans. Graph.},
  volume={35},
  number={1},
  pages={1--1},
  year={2015}
}

@article{khungurn2017fast,
  title={Fast rendering of fabric micro-appearance models under directional and spherical gaussian lights},
  author={Khungurn, Pramook and Wu, Rundong and Noeckel, James and Marschner, Steve and Bala, Kavita},
  journal={ACM Transactions on Graphics (TOG)},
  volume={36},
  number={6},
  pages={1--15},
  year={2017},
  publisher={ACM New York, NY, USA}
}

@article{zhao2013modular,
  title={Modular flux transfer: efficient rendering of high-resolution volumes with repeated structures},
  author={Zhao, Shuang and Ha{\v{s}}an, Milo{\v{s}} and Ramamoorthi, Ravi and Bala, Kavita},
  journal={ACM Transactions on Graphics (TOG)},
  volume={32},
  number={4},
  pages={1--12},
  year={2013},
  publisher={ACM New York, NY, USA}
}

@inproceedings{luan2017fiber,
  title={Fiber-Level On-the-Fly Procedural Textiles},
  author={Luan, Fujun and Zhao, Shuang and Bala, Kavita},
  booktitle={Computer Graphics Forum},
  volume={36},
  number={4},
  pages={123--135},
  year={2017},
  organization={Wiley Online Library}
}

@inproceedings{schroder2011volumetric,
  title={A volumetric approach to predictive rendering of fabrics},
  author={Schroder, K and Klein, Reinhard and Zinke, Arno},
  booktitle={Computer Graphics Forum},
  volume={30},
  number={4},
  pages={1277--1286},
  year={2011},
  organization={Wiley Online Library}
}

@inproceedings{jin2022woven,
  title={Woven Fabric Capture from a Single Photo},
  author={Jin, Wenhua and Wang, Beibei and Hasan, Milos and Guo, Yu and Marschner, Steve and Yan, Ling-Qi},
  booktitle={SIGGRAPH Asia 2022 Conference Papers},
  pages={1--8},
  year={2022}
}

@article{irawan2012specular,
  title={Specular reflection from woven cloth},
  author={Irawan, Piti and Marschner, Steve},
  journal={ACM Transactions on Graphics (TOG)},
  volume={31},
  number={1},
  pages={1--20},
  year={2012},
  publisher={ACM New York, NY, USA}
}

@article{sadeghi2013practical,
  title={A practical microcylinder appearance model for cloth rendering},
  author={Sadeghi, Iman and Bisker, Oleg and De Deken, Joachim and Jensen, Henrik Wann},
  journal={ACM Transactions on Graphics (TOG)},
  volume={32},
  number={2},
  pages={1--12},
  year={2013},
  publisher={ACM New York, NY, USA}
}

@article {schlick1994,
journal = {Computer Graphics Forum},
title = {{An Inexpensive BRDF Model for Physically-based Rendering}},
author = {Schlick, Christophe},
year = {1994},
publisher = {Blackwell Science Ltd and the Eurographics Association},
ISSN = {1467-8659},
DOI = {10.1111/1467-8659.1330233}
}

@article{sztrajman2021nbrdf,
author={Alejandro Sztrajman and Gilles Rainer and Tobias Ritschel and Tim Weyrich},
title = {Neural BRDF Representation and Importance Sampling},
journal = {Computer Graphics Forum},
year = {2021},
doi = {https://doi.org/10.1111/cgf.14335},
url = {https://onlinelibrary.wiley.com/doi/abs/10.1111/cgf.14335}
}

@article{leaf2018stanfordyarn,
author = {Leaf, Jonathan and Wu, Rundong and Schweickart, Eston and James, Doug L. and Marschner, Steve},
title = {Interactive Design of Periodic Yarn-Level Cloth Patterns},
year = {2018},
issue_date = {December 2018},
publisher = {Association for Computing Machinery},
address = {New York, NY, USA},
volume = {37},
number = {6},
issn = {0730-0301},
url = {https://doi.org/10.1145/3272127.3275105},
doi = {10.1145/3272127.3275105},
journal = {ACM Trans. Graph.},
month = {dec},
articleno = {202},
numpages = {15}
}

@ARTICLE{chen2020ibrdf,
author={Chen, Zhe and Nobuhara, Shohei and Nishino, Ko},
journal={IEEE Transactions on Pattern Analysis and Machine Intelligence}, 
title={Invertible Neural BRDF for Object Inverse Rendering}, 
year={2022},
volume={44},
number={12},
pages={9380-9395},
doi={10.1109/TPAMI.2021.3129537}
}

@INPROCEEDINGS{he2015prelu,
  author={He, Kaiming and Zhang, Xiangyu and Ren, Shaoqing and Sun, Jian},
  booktitle={2015 IEEE International Conference on Computer Vision (ICCV)}, 
  title={Delving Deep into Rectifiers: Surpassing Human-Level Performance on ImageNet Classification}, 
  year={2015},
  volume={},
  number={},
  pages={1026-1034},
  doi={10.1109/ICCV.2015.123}
}

@inproceedings{nair2010relu,
author = {Nair, Vinod and Hinton, Geoffrey E.},
title = {Rectified Linear Units Improve Restricted Boltzmann Machines},
year = {2010},
isbn = {9781605589077},
publisher = {Omnipress},
address = {Madison, WI, USA},
booktitle = {Proceedings of the 27th International Conference on International Conference on Machine Learning},
pages = {807–814},
numpages = {8},
location = {Haifa, Israel},
series = {ICML'10}
}

@inproceedings{zhu2023realistic,
  title={A Realistic Surface-based Cloth Rendering Model},
  author={Zhu, Junqiu and Jarabo, Adrian and Aliaga, Carlos and Yan, Ling-Qi and Chiang, Matt Jen-Yuan},
  booktitle={ACM SIGGRAPH 2023 Conference Proceedings},
  pages={1--9},
  year={2023}
}

@article{kuznetsov2021neumip,
  title={NeuMIP: multi-resolution neural materials},
  author={Kuznetsov, Alexandr and Mullia, Krishna and Xu, Zexiang and Ha{\v{s}}an, Milo{\v{s}} and Ramamoorthi, Ravi},
  journal={ACM Transactions on Graphics (TOG)},
  volume={40},
  number={4},
  pages={1--13},
  year={2021},
  publisher={ACM New York, NY, USA}
}

@inproceedings{rainer2019neural,
  title={Neural BTF compression and interpolation},
  author={Rainer, Gilles and Jakob, Wenzel and Ghosh, Abhijeet and Weyrich, Tim},
  booktitle={Computer Graphics Forum},
  volume={38},
  number={2},
  pages={235--244},
  year={2019},
  organization={Wiley Online Library}
}

@inproceedings{sattler2003efficient,
  title={Efficient and realistic visualization of cloth},
  author={Sattler, Mirko and Sarlette, Ralf and Klein, Reinhard},
  booktitle={Rendering techniques},
  pages={167--178},
  year={2003}
}

@inproceedings{aliaga2017appearance,
  title={An appearance model for textile fibers},
  author={Aliaga, Carlos and Castillo, Carlos and Gutierrez, Diego and Otaduy, Miguel A and Lopez-Moreno, Jorge and Jarabo, Adrian},
  booktitle={Computer Graphics Forum},
  volume={36},
  number={4},
  pages={35--45},
  year={2017},
  organization={Wiley Online Library}
}

@inproceedings{xu2023neusample,
  title={NeuSample: Importance Sampling for Neural Materials},
  author={Xu, Bing and Wu, Liwen and Hasan, Milos and Luan, Fujun and Georgiev, Iliyan and Xu, Zexiang and Ramamoorthi, Ravi},
  booktitle={ACM SIGGRAPH 2023 Conference Proceedings},
  pages={1--10},
  year={2023}
}

%-------------------------------------------------------------------------
\section*{Appendix A: Derivation of Rotation Angle.}

Borrowing concepts from rotational dynamics, the tangential velocity, $v$, on the surface of the yarn can be defined as:
\begin{equation}
    v = R\omega
    \label{eq:v}
\end{equation}
where $R$ represents the radius of the yarn and $\omega$ denotes the rotational velocity. The rotational velocity can be expressed as:
\begin{equation}
    \omega = \frac{\Delta \theta}{\Delta l}
    \label{eq:omega}
\end{equation}
where $\Delta \theta$ is the change in angular position and $\Delta l$ is change in position along the yarn tangent. From Equation \ref{eq:alpha} and \ref{eq:omega}, we can obtain:
\begin{equation}
    \alpha = \frac{2R\Delta n}{\Delta l} = \frac{R \Delta \theta}{\pi \Delta l} = \frac{R}{\pi} \omega
\end{equation}
Substituting the previous equation with \ref{eq:v}, we obtain:
\begin{equation}
    v = \pi \alpha
\end{equation}
Given that $v$ represents the rate of change of the bi-tangent with respect to the yarn tangent, and considering that they are perpendicular to each other, forming a right-angle triangle, we deduce:
\begin{equation}
\begin{split}
    \tan \tilde{\phi} &= v \\
    \tilde{\phi} &= \tan^{-1}(\pi \alpha)
\end{split}
\label{eq:phi}
\end{equation}

\end{document}